\documentclass[prb,aps,twocolumn]{revtex4-1}
\usepackage{latexsym}
\usepackage{amssymb}
\usepackage{amsmath}
\usepackage{amsbsy}
\usepackage{mathtools}
\usepackage{mathrsfs}
\usepackage[pdftex]{graphicx}
\usepackage{epstopdf}
\usepackage{color}
\usepackage{soul}
\usepackage[english]{babel}

\begin{document}

\title{Electron-nuclear spin dynamics of
Ga$^{2+}$ paramagnetic centers probed by spin
dependent recombination: A master equation approach}

\author{V. G. Ibarra-Sierra$^1$}\author{J. C. Sandoval-Santana$^1$}
\author{S. Azaizia$^2$}\author{H. Carr\`ere$^2$}
\author{L.A. Bakaleinikov$^3$}
\author{V. K. Kalevich$^3$}
\author{E. L. Ivchenko$^3$}
\author{X. Marie$^2$}\author{T. Amand$^2$}
\author{A. Balocchi$^2$}
\author{A. Kunold$^4$}
\address{
$^1$Departamento de F\'isica, Universidad Aut\'onoma Metropolitana
Iztapalapa, Av. San Rafael Atlixco 186, Col. 
Vicentina, 09340 Cuidad de M\'exico, M\'exico\\
$^2$Universit\`e de Toulouse, INSA-CNRS-UPS, LPCNO, 135 avenue de Rangueil,
 31077 Toulouse, France\\
$^3$Ioffe  Physical-Technical Institute, 194021 St. Petersburg, Russia\\
$^4$\'Area de F\'isica Te\'orica y Materia Condensada,
Universidad Aut\'onoma Metropolitana  Azcapotzalco,
Av. San Pablo 180, Col. Reynosa-Tamaulipas,
02200 Cuidad de M\'exico, M\'exico}

\date{\today}
\begin{abstract}
Similar to nitrogen-vacancy centers in diamond and
impurity atoms in silicon, interstitial gallium deep paramagnetic centers in
GaAsN have been proven to have useful characteristics
for the development of spintronic devices.
Among other interesting properties, under circularly polarized light,
gallium centers act as spin filters that dynamically
polarize free and bound electrons reaching
record  spin polarizations (close to 100\%).
Furthermore, the recent observation of the amplification
of the spin filtering effect under a Faraday configuration magnetic
field has suggested
that the hyperfine interaction that couples
bound electrons and nuclei
permits the optical manipulation of its nuclear spin polarization.
Even though the mechanisms behind the nuclear spin polarization
in gallium centers are fairly well understood, the origin of nuclear spin
relaxation and the formation of an Overhauser-like magnetic field
remain elusive.
In this work we develop a model based on the master equation approach
to describe the evolution of electronic and nuclear spin polarizations of gallium
centers interacting with free electrons and holes.
Our results are in good agreement with existing
experimental observations. In particular, we are able
to reproduce
the amplification of the spin filtering effect under
a circularly polarized excitation 
in a Faraday configuration magnetic field. 
In regard to the nuclear spin relaxation,  the roles of nuclear
dipolar and quadrupolar interactions
are discussed. Our findings show that, besides the hyperfine interaction,
the spin relaxation mechanisms are key  to understand
the amplification of the spin filtering effect and the appearance
of the Overhauser-like magnetic field.
To gain a deeper insight in the interplay of the hyperfine
interaction and the relaxation mechanisms,
we have also performed calculations in the pulsed excitation regime.
Based on our model's results
we propose an experimental protocol based on time resolved spectroscopy.
It consists of a pump-probe photoluminescence scheme
that would allow the detection and the tracing of the electron-nucleus
flip-flops through time resolved PL measurements.
\end{abstract}

\maketitle


\section{Introduction}

Negatively charged nitrogen-vacancy centers in diamond\cite{PhysRevB.93.035402,Sar2012,PhysRevB.80.041201,PhysRevB.78.094303,PhysRevLett.113.246801},
phosphorous atom impurities in silicon\cite{kane,Pla2012,PhysRevB.71.014401,
:/content/aip/journal/apl/104/9/10.1063/1.4867905,PhysRevB.74.195301,PhysRevB.92.115206}
 and other schemes based
on point defects embedded in semiconductors
have been widely studied as alternatives
to develop quantum bits\cite{Ladd2010,Wrachtrup2016}.
One of the necessary conditions for quantum computing is
long electron spin decoherence times to insure a minimum of fault-tolerance
 \cite{PhysRevLett.105.187602,PhysRevB.74.195301}.
In diamond's nitrogen-vacancy centers\cite{Hanson352}, silicon vacancies in silicon carbide\cite{PhysRevX.6.031014}, silicon\cite{Pla2013}
and any III-V based quantum dots\cite{PhysRevLett.94.116601,PhysRevB.74.195301}
the fluctuating nuclear bath is the main source limiting spin coherence time.
The nuclear dipole-dipole
interaction is believed to be the dominant mechanism
behind the diffusion-induced electron-spin decoherence\cite{PhysRevB.82.121201}.
To protect the dynamics of the nuclear spins of point defects  
from the decoherence induced by the environment, semiconductors
mainly composed of spin-zero isotopes as silicon and carbon
are preferred over III-V semiconductors\cite{PhysRevLett.113.246801}.
Even though the two stable isotopes of Ga, $^{69}$Ga and $^{71}$Ga,
have nuclear spin $3/2$,
in dilute nitride GaAsN, point interstitial defects give rise to paramagnetic
centers that have very peculiar and useful properties.
One of them is the spin dependent recombination (SDR)
\cite{Kalevich2005,Kalevich2007,:/content/aip/journal/apl/95/24/10.1063/1.3275703,
:/content/aip/journal/apl/95/4/10.1063/1.3186076,
:/content/aip/journal/apl/95/24/10.1063/1.3273393,0953-8984-21-17-174211,
:/content/aip/journal/apl/96/5/10.1063/1.3299015,
:/content/aip/journal/apl/96/5/10.1063/1.3299015,0953-8984-22-46-465804}.
In Ga(In)NAs alloys, Ga$^{+2}_i$ paramagnetic centers
with only one bound electron can selectively capture another
conduction band (CB) electron
with the opposite spin orientation
\cite{PhysRevB.30.931,
:/content/aip/journal/apl/96/5/10.1063/1.3299015,0953-8984-22-46-465804,
WEISBUCH1974141,:/content/aip/journal/apl/103/5/10.1063/1.4816970}.
Due to this mechanism, paramagnetic centers act as a spin
filter that blocks the recombination of CB electrons with the same spin
and efficiently capture electrons whose spin is in the opposite direction.
In the centers, the bound and captured electron form a singlet state
that is destroyed as either one of the electrons recombines to the valence
band (VB). It is important to note that while the
lifetime of a CB electron
with the opposite spin to the paramagnetic electrons
is a few pico-seconds, the lifetime of a spin polarized CB electron
when the majority of bound electrons are polarized parallel to it may extend to nanoseconds.
As a consequence
CB electrons spin polarization can reach over 80\%
under circularly polarized incident light.
Additionally the photoluminescence (PL) intensities can be as high
as 800\% under circularly polarized optical excitation compared to a linearly polarized
one  \cite{Kalevich2007,:/content/aip/journal/apl/95/24/10.1063/1.3275703}.
The increase in CB electron population allows even for
the detection of electron spin polarization by
electrical means due to a giant photoconductivity effect under circulalrly polarized
light\cite{:/content/aip/journal/apl/95/24/10.1063/1.3273393,PhysRevB.83.165202}.

Whereas in diamond and silicon the optical excitation
acts directly on the point defects, in GaAsN
the bound electron is dynamically spin polarized due
to the recombination of spin polarized CB electrons
on the paramagnetic centers.
Although this mechanism is radically different, Ga$^{2+}_i$ centers themselves
are very similar to nitrogen vacancies and silicon phosphorous atoms.
Recent experiments on GaAsN  subject to a weak magnetic field in Faraday configuration
\cite{PhysRevB.85.035205,Kalevich2013,PhysRevB.87.125202,Puttisong2013}
have shown consistently an enhancement in the spin-filtering mechanism
in comparison to the zero magnetic field case.
The general agreement among different
models\cite{PhysRevB.30.931,Puttisong2013,PhysRevB.90.115205,
PhysRevB.91.205202}
is that the hyperfine interaction (HFI)
between the bound electron and the nucleus in the centers 
is the key element behind this phenomenon.
At low magnetic fields, the spin-filtering effect is reduced due to spin state mixing
induced by the HFI. For higher values of the magnetic field such that the Zeeman
energy exceeds the HFI, the pure bound electron spin states and, consequently, the
spin filtering effect are recovered.
Even though the role of the HFI is well established, some aspects of these phenomena
lacks full understanding.
Some observations point to the existence of an Overhauser-like effective
magnetic field\cite{PhysRevB.85.035205,Kalevich2013} whose origin is yet unclear. It manifests as a shift in the band to band PL intensity
or the degree of CB electron spin polarization
as functions of the magnetic field. Both features shift to the positive
and negative regions of the magnetic field for right  and
left circularly polarized light respectively.
Another aspect that needs further study is the nuclear interaction
between Ga$^{2+}_i$ centers and adjacent Ga atoms that would lead to nuclear spin relaxation (NSR).
The Overhauser-like magnetic field has been correctly reproduced in Ref. [\cite{PhysRevB.90.115205}],
however, nuclear spins in this model are assumed to relax very rapidly and the origin of the relaxation
mechanism is unclear.
On the other hand, the model presented in Ref. [\cite{PhysRevB.91.205202}]
considers two phenomenological and arbitrary NSR times
for traps with one and two bound electrons.
Despite the improvements in this work,
the model considers Ga centers with $1/2$ nuclear spins
instead of $3/2$ in order to simplify the kinetic equations.

In this paper we examine the spin dynamics
in GaAsN alloys.
We propose a model
based on the master equation for the density matrix
that describes the main interactions between
CB electrons, VB holes and paramagnetic traps.
It addresses the problems
on spin relaxation mechanisms and Overhauser-like magnetic field.
In fact, our results show that the Overhauser-like magnetic field
strongly depends on the NSR times and
the mechanisms behind them.

The model developed here is based on
the preexisting two charge model (TCM)
\cite{Kalevich2005,Kalevich2007,0953-8984-22-46-465804,WEISBUCH1974141}.
In addition to the SDR
processes, it contains
the mechanisms that give rise so
NSR and HFI.
NSR is addressed through the
Wangsness, Bloch, and Redfield relaxation theory
\cite{ PhysRev.89.728,REDFIELD19651,PhysRev.142.179,kowalewski2006nuclear}.
Two interactions are explored as possible candidates
to produce the NSR: dipolar interaction between neighbouring
nuclei\cite{PhysRev.89.728} and quadrupolar
\cite{PhysRev.89.728,PhysRevLett.109.166605}
interaction with charge fluctuations in the environment.

To further explore the role of HFI in Ga centers,
we have studied the time resolved electronic and
nuclear spin polarizations under pulsed excitation.
Using these results we outline a method
based on a pump-probe photoluminescence scheme
to trace the coherent evolution of coupled electrons
and nuclear spins as they flip-flop due to the HFI.

This paper is organized as follows. In Sec. \ref{model}
we introduce the master equation model that
describes the key processes in the spin
dynamics of electrons and nuclei in GaAsN.
The role of the HFI and the NSR mechanism
are discussed in this section.
The mathematical forms of
the dipolar and quadrupolar dissipators are
introduced in Sec. \ref{nsrsec}.
In Sec. \ref{nsrcwregime} we establish
the main mechanism behind the nuclear
spin relaxation in Ga$^{2+}$ centers by
comparing the theoretical
results issued by the model with
previously existing experimental results.
Simulations in the pulsed excitation regime
are presented in Sec. \ref{coherent}.
In Sec. \ref{summ} we summarize the main
results.

\section{Master equation}\label{model}
The system mainly consists of four elements: VB holes, CB electrons,
unpaired traps (UT) and
paired traps (PT). Whereas VB holes' spin relaxes with a characteristic
times below $1\,\mathrm{ps}$\cite{PhysRevLett.89.146601},
CB electronic spin relaxes on a typical time in the range $100-400\, \mathrm{ps}$\cite{0953-8984-21-17-174211,kalevich2009hanle}. Therefore, for the
sake of simplicity, we consider VB holes to be unpolarized and
electrons may occupy spin-up or
spin-down states.
Gallium paramagnetic traps or UTs can be understood as a $3/2$
nuclear spin coupled to a $1/2$ bound electron spin via HFI.
On the other hand, when a PT is formed if a CB electron
is captured by an UT, the bound and captured electrons form a singlet state
that cannot interact with the nuclear spin.
Thus the quantum state basis that describes
this system must have:
(i) one state for holes in the VB, 
(ii) two states for the spin up and spin down CB electrons,
(iii) eight states that account for the nucleus-bound
electron system in the UTs and
(iv) four states for the nuclear spin in the PTs.
The complete  quantum state basis is therefore
given by
\begin{multline}\label{blocks}
  \mathbf{
  \mathcal{B}} =
  \bigg\{
  \vert 1\rangle =\vert h\rangle, 
  \vert 2\rangle=\vert\downarrow\rangle, 
  \vert 3\rangle=\vert\uparrow\rangle,
  \vert 4\rangle=\vert -\frac{3}{2},\downarrow\rangle,\\
  \vert 5\rangle=\vert -\frac{1}{2},\downarrow\rangle,
  \vert 6\rangle=\vert \frac{1}{2},\downarrow\rangle,
  \vert 7\rangle=\vert \frac{3}{2},\downarrow\rangle,\\
  \vert 8\rangle=\vert -\frac{3}{2},\uparrow\rangle,
  \vert 9\rangle=\vert -\frac{1}{2},\uparrow\rangle,
  \vert 10\rangle=\vert \frac{1}{2},\uparrow\rangle,\\
  \vert 11\rangle=\vert\frac{3}{2},\uparrow\rangle,
  \vert 12\rangle=\vert-\frac{3}{2},\uparrow\downarrow\rangle,\\
  \vert 13\rangle=\vert-\frac{1}{2},\uparrow\downarrow\rangle,
  \vert 14\rangle=\vert\frac{1}{2},\uparrow\downarrow\rangle,
  \vert 15\rangle=\vert\frac{3}{2},\uparrow\downarrow\rangle
  \bigg\},
\end{multline}
where $\vert h\rangle$ is the VB hole state and, $\vert\uparrow\rangle$ and
$\vert\downarrow\rangle$ are the spin-up and spin-down CB electron states
respectively. The following eight states $\vert -\frac{3}{2},\downarrow\rangle$,
$\vert -\frac{1}{2},\downarrow\rangle$, $\dots$, $\vert\frac{3}{2},\uparrow\rangle$
are the bound-electron and nuclear spin states projected along the $z$ axis
corresponding to the UT.
Finally, the PTs are described by the nuclear-spin states
$\vert-\frac{3}{2},\uparrow\downarrow\rangle$, $\dots$,
$\vert\frac{3}{2},\uparrow\downarrow\rangle$.

The dynamics of the four parts of the system and their interactions can be described
through the master equation
\begin{equation}\label{masterequation}
 \frac{d\hat{\rho}}{dt}=\frac{i}{\hbar}\left[\hat \rho,\hat H\right]+\mathcal{D}\left(\hat\rho\right),
\end{equation}
where $\hat\rho$ is the density matrix, $\hat H$ is the Hamiltonian and $\mathcal{D}\left(\hat\rho\right)$
is the dissipator. The Hamiltonian contains the internal
interactions among the four components of the system.
We are interested in the combined effect of an external magnetic field and the HFI in Ga centers,
therefore the Hamiltonian must contain Zeeman and HFI terms.
The interactions among the different parts of the system and the surroundings are accounted for
by the Dissipator.
These are mostly interactions with the electromagnetic field, occurring during recombination or
excitation processes, or interactions with the nuclear spin environment.
The main processes introduced in our
model are schematized in Fig. 1. 

\subsection{Hamiltonian: Zeeman and hyperfine interactions}
The Hamiltonian is given by
  \begin{eqnarray}\label{spinHamiltonian}
    \hat H &=& \hbar\boldsymbol{\omega}\cdot \hat{\boldsymbol{S}}
            +\hbar\boldsymbol{\Omega}\cdot\hat{\boldsymbol{S}}_c
            +A\hat{\boldsymbol{I}}_1\cdot\hat{\boldsymbol{S}}_c.
  \end{eqnarray}
The first and second terms in the Hamiltonian correspond to the Zeeman interaction for CB
and bound electrons in Ga centers respectively. 
In these terms $\boldsymbol{\omega}=g\mu_{B}\boldsymbol{B}/\hbar$,
$\boldsymbol{\Omega}=g_{c}\mu_{B}\boldsymbol{B}/\hbar$, $\boldsymbol{B}$ is the
external magnetic field, $\mu_B$ is the Bohr magneton  and, $g$ and $g_c$ are
the CB and bound electrons gyromagnetic factors.
The HFI term, the third one on the right-hand side of Eq. (\ref{spinHamiltonian}), 
couples  the bound electron and the nuclear spin
in UTs. The hyperfine parameter is given by $A$.

The spin operator for CB electrons that appears in the first Zeeman term of the Hamiltonian is
$\hat{\boldsymbol{S}}=(\hat{S}_{x},\hat{S}_{y},\hat{S}_{z})$.
In UTs, the bound electron spin and nuclear spin operators entering the HFI term are
$\hat{\boldsymbol{S}}_c=(\hat{S}_{cx},\hat{S}_{cy},\hat{S}_{cz})$,
and $\hat{\boldsymbol{I}}_{1}=(\hat{I}_{1x},\hat{I}_{1y},\hat{I}_{1z})$ respectively.
As the singlet state formed in the PTs  interacts neither with the external magnetic field
nor with their nuclear spin
$\hat{\boldsymbol{I}}_{2}=(\hat{I}_{2x},\hat{I}_{2y},\hat{I}_{2z})$, it does
not appear in the Hamiltonian.

\subsection{Density matrix operator space}
The relaxation mechanisms of CB electrons and nuclei
in the Ga centers are described by the dissipator $\mathcal{D}(\hat{\rho})$. Also, the photoexcitation and  recombination of electrons will be accounted for by $\mathcal{D}$.

Our approach to formulating a suitable dissipator consists
in expanding the  relevant operators as linear combination
of the elements of an operator vector space.
In principle this set should be formed  in the basis (1) by linearly independent
$15\times 15$ Hermitian matrices.
However, the density matrix structure is considerably
simplified by assuming that the four components of the system  (CB, UT, PT, VB) are interconnected 
only by the dissipator. 
As the four parts of the system are exclusively coupled by the recombination or excitation processes, this is a reasonable assumption. 
Thus, the density matrix operator can be presented in the block diagonal form:
\begin{equation}
\hat{\rho}=\left(
\begin{array}{ccccc}
\hat{\rho}_{\text{\tiny VB}} & & & & \\
  & \hat{\rho}_{\text{\tiny CB}} &  \\
  & &  \hat{\rho}_{1} & \\
  & & & \hat{\rho}_{2} \\
\end{array}
\right)
\end{equation}
where the four blocks $\hat{\rho}_{\text{\tiny VB}}$ ($1\times 1$),
$\hat{\rho}_{\text{\tiny CB}}$ ($2\times 2$),
$\hat{\rho}_{1}$ ($8\times 8$)
 and $\hat{\rho}_{2}$ ($4\times 4$)
are the partial density matrices of VB holes, CB electrons, UTs and PTs
respectively.
Given that $\hat\rho$ takes the form of a block diagonal matrix 
since no coherences can arise between the four components,
it suffices to consider the smaller vector space 
of $85$ Hermitian matrices
that generate the four blocks.

We start by finding an internal space of Hermitian matrices
\begin{equation}\label{or:lambda}
 \Lambda=\left\{\hat \lambda_1,\hat \lambda_2,\dots,\hat \lambda_{n}\right\},
\end{equation}
that spans the $n=85$ relevant elements of the $15\times 15$ matrix. 
The generators in this set can be chosen in such a way that they are Hermitian 
and orthogonal with respect to the scalar product given by the trace
\begin{equation}\label{producto-interno}
 \langle\hat{\lambda}_{i},\hat{\lambda}_{j}\rangle\equiv \text{Tr}\left[\hat \lambda_{i}^{\dagger} \hat 
 \lambda_{j}\right]=\delta_{i,j}\text{Tr}\left[\hat\lambda_{i}^{2}\right].
\end{equation}
This choice conveniently links the inner product with the expected value
of a given operator $\hat O$
\begin{equation}
O =\mathrm{Tr}\left[\hat O \hat \rho\right]= \langle \hat\rho, \hat O\rangle,
\end{equation}
acting on $\hat{\rho}$.
In this manner any operator can be expanded as
a linear combination of the elements of 
(\ref{or:lambda}) as
\begin{equation}\label{any:operator}
 \hat O=\sum_{q=1}^{85}
 \frac{\text{Tr}\left[\hat\lambda_q\hat O\right]}{\text{Tr}\left[\hat\lambda_q^2\right]}
 \hat \lambda_q.
\end{equation}

A very convenient set of operators is the one formed
by the generators of the unitary groups $U(1)$ (VB holes),
$U(2)$  (CB electrons), $U(4)$ (PTs) and $U(8)$ (UTs).
The operators forming this set are not only of physical significance
but they are also linearly independent and
orthogonal with respect to the trace.
Explicitly,  the set of operators in (\ref{or:lambda}) is given by
\begin{equation}\label{explicit:lambda}
 \Lambda=\left\{\hat p,\hat{S}_{i},\hat{U}_{k,j,i},\hat{T}_{j,i}\right\},
\,\,\,\, i,j,k=0,1,2,3;
\end{equation}
where $\hat p$, $\hat S_i$, $\hat{U}_{k,j,i}$ and $\hat{T}_{j,i}$
generate the VB, CB, UT and PT bolcks.
The VB hole population density operator can be represented
by the matrix
\begin{equation}
\hat{p}=\left(
\begin{array}{cccc}
1 & & & \\
  & \hat{0}_{2\times 2} & &  \\
  &  &  \hat{0}_{8\times 8} &  \\
  & & & \hat{0}_{4\times 4}  \\
\end{array}
\right).
\end{equation}
The operators that generate the
the CB block can be compiled
in the matrix
\begin{equation}
\hat{S}_{i}=\left(
\begin{array}{ccccc}
0 & & & & \\
  & \hat s_i &  \\
  & &  \hat{0}_{8\times 8} & \\
  & & & \hat{0}_{4\times 4} \\
\end{array}
\right),\,\,\,\, i=0,1,2,3,
\end{equation}
where the electron population density in the CB is given by $\hat n=\hat S_0$
and their spin operators are $\hat S_i$ for $i=1,2,3$.
Here $\hat s_0=\hat 1_{2\times 2}$ is the $2\times 2$
identity matrix,
and $\hat s_i$ for $i=1,2,3$ are the standard Pauli spin
matrices that fullfil the usual spin commutation relations
\begin{equation}
\left[\hat s_i,\hat s_j\right]=i \sum_{k=1,2,3}\epsilon_{ijk}\hat s_k.
\end{equation}
This definition allows us to write the
matrices that generate the UT and PT blocks
in the compact forms
\begin{equation}
\hat{U}_{k,j,i}=\left(
\begin{array}{ccccc}
0 & & & & \\
  & \hat{0}_{2\times 2} &  \\
  & &  \hat{s}_{k}\otimes\hat{s}_{j}\otimes\hat{s}_{i} \\
  & & & \hat{0}_{4\times 4}\\
\end{array}
\right),
\end{equation}
and
\begin{equation}
\hat{T}_{j,i}=\left(
\begin{array}{ccccc}
0 & & & & \\
  & \hat{0}_{2\times 2} &  \\
  & &  \hat{0}_{8\times 8} & \\
  & & & \hat{s}_{j}\otimes\hat{s}_{i} \\
\end{array}
\right).
\end{equation}
According to this scheme the population density of
UTs is $\hat{N}_{1}=\hat{U}_{0,0,0}$,
and the one for  PTs is $\hat{N}_{2}=\hat{T}_{0,0}$.
Similarly,  the CB electrons' spin operators are: $\hat{S}_{x}=\hat{S}_{1}$,  
$\hat{S}_{y}=\hat{S}_{2}$ and $\hat{S}_{z}=\hat{S}_{3}$.
We have the same case for the bound electron spin
operator components in UTs  where  $\hat{S}_{cx}= \hat{U}_{1,0,0}$,
$\hat{S}_{cy}= \hat{U}_{2,0,0}$ and $\hat{S}_{cz}=\hat{U}_{3,0,0}$. 
The operators of the nuclear spin of UTs and PTs
can be expressed as linear combinations of the elements
of $\Lambda$ as
\begin{eqnarray}
 \hat{\boldsymbol{I}}_{1}&=&\mathcal{M}\hat{\boldsymbol{\mathcal{U}}},\\
 \hat{\boldsymbol{I}}_{2}&=&\mathcal{M}\hat{\boldsymbol{\mathcal{T}}},
\end{eqnarray}
where 
\begin{eqnarray}
\hat{\boldsymbol{\mathcal{U}}}^{\top}&=&(\hat{U}_{0,0,1},\hat{U}_{0,1,1},
\hat{U}_{0,2,2},\hat{U}_{0,0,2},\hat{U}_{0,1,2}, 
\hat{U}_{0,2,1},\nonumber \\
&&\hat{U}_{0,0,3},\hat{U}_{0,3,0}),\\
\hat{\boldsymbol{\mathcal{T}}}^{\top}&=&(\hat{T}_{0,1},\hat{T}_{1,1} ,\hat{T}_{2,2},\hat{T}_{0,2},\hat{T}_{1,2}, 
\hat{T}_{2,1},\hat{T}_{0,3},\hat{T}_{3,0}),
\end{eqnarray}
and
\begin{equation}
\mathcal{M}=
\left[
\begin{array}{cccccccc}
\sqrt{3} & 2 & 2 & 0 & 0 & 0 & 0 & 0\\
0 & 0 & 0 & \sqrt{3} & -2 & 2 & 0 & 0\\
0 & 0 & 0 & 0 & 0 & 0 & 1 & 2\\
\end{array}
\right].
\end{equation}

\subsection{Dissipator}

The dissipator can be separated in six parts
\begin{equation}\label{general:dissipator}
\mathcal{D}\left(\hat\rho\right)=\hat{\mathcal{G}}
+\hat{\mathcal{D}}_{\mathrm{SDR}}
+\hat{\mathcal{D}}_{\mathrm{S}}+\hat{\mathcal{D}}_{\mathrm{SC}}
+\hat{\mathcal{D}}_{1}+\hat{\mathcal{D}}_{2}.
\end{equation}
Here $\hat{\mathcal{G}}$ contains the VB hole and CB electron photogeneration terms.
The SDR processes that mainly consist of the selective capture of CB electrons in UTs
according to their relative spin orientation and the subsequent
recombination to the VB are described by the $\hat{\mathcal{D}}_{\mathrm{SDR}}$ dissipator.
CB and bound electron spin relaxation are accounted by
$\hat{\mathcal{D}}_{\mathrm{S}}$ and
$\hat{\mathcal{D}}_{\mathrm{SC}}$.
NSR in UT's and PT's is introduced through the dissipators
$\hat{\mathcal{D}}_{1}$ and $\hat{\mathcal{D}}_{2}$.

The term $\hat{\mathcal{G}}$ that models the generation of electrons 
is given by
\begin{equation}
 \hat{\mathcal{G}}=\left(G_{\uparrow}+G_{\downarrow}\right)\left(\hat{p}+\hat{n}\right)
 +2\left(G_{\uparrow}-G_{\downarrow}\right)\hat{\boldsymbol{e}}\cdot\hat{\boldsymbol{S}},
\end{equation}
where $G_{\uparrow}$ and $G_{\downarrow}$ are the
spin up and spin down electron generation rates. The unitary vector $\boldsymbol{e}$ points
in the direction of the incident light.

To build the $\mathcal{D}_{\mathrm{SDR}}$ part of the dissipator
we resort to the TCM\cite{Kalevich2005,Kalevich2007,0953-8984-22-46-465804,WEISBUCH1974141}
given by the following kinetic equations ($\hbar$=1)
\begin{align}
\dot{n}
  &=-c_n\left(nN_1-4\boldsymbol{S}\cdot\boldsymbol{S}_c\right)
  +G_{\uparrow}+G_{\downarrow},
  \label{mag:eq3}\\
\dot{p}&=-c_pN_2p+G_{\uparrow}+G_{\downarrow},\\
\dot{N_1}
  &=-c_n\left(nN_1-4\boldsymbol{S}\cdot\boldsymbol{S}_c\right)
  +c_pN_2p,\\
\dot{N_2}
  &=c_n\left(nN_1-4\boldsymbol{S}\cdot\boldsymbol{S}_c\right)
  -c_pN_2p.\label{mag:eq6}\\
\dot{\boldsymbol{S}}
  &=-c_n\left(\boldsymbol{S}N_1-\boldsymbol{S}_cn\right)
  -\frac{1}{\tau_s}\boldsymbol{S}+\boldsymbol{\omega}\times\boldsymbol{S}
  \nonumber \\
  &\,\,\,\,\,\,\,\,\,\,\,\,\,\,\,\,\,\,\,\,\,\,\,\,
  +\frac{G_{\uparrow}-G_{\downarrow}}{2}\hat{\boldsymbol{e}},\label{mag:eq1}\\
\dot{\boldsymbol{S}}_c
  &=-c_n\left(\boldsymbol{S}_cn-\boldsymbol{S}N_1\right)
  -\frac{1}{\tau_{sc}}\boldsymbol{S}_c
  +\boldsymbol{\Omega}\times\boldsymbol{S}_c.\label{mag:eq2}
\end{align}
In the previous equations, the population densities of CB electrons and VB holes are
given by $n$ and $p$ respectively.
The density of UTs is  $N_1$, $N_2$ is the density
of electron singlets hosted
by the centers (PTs), and consequently $N_c=N_1+N_2$ is
the total density of Ga centers.
The vectors $\boldsymbol{S}$ and $\boldsymbol{S}_c$
represent the average of free and bound electron spin polarizations.
The spin dependent capture of electrons in the Ga centers
is ensured by the recombination rate terms
$c_n\left(nN_1-4\boldsymbol{S}\cdot\boldsymbol{S}_c\right)$
and $c_n\left(\boldsymbol{S}N_1-\boldsymbol{S}_cn\right)$
where $c_n$ is a constant. Notice that these two terms vanish
when the system is fully polarized, i.e.
$S_{z}=n/2$, $S_{x}=S_{y}=0$, $S_{cz}=N_1/2$
and $S_{cx}=S_{cy}=0$. The recombination rate of one of the
electrons trapped in the Ga centers to the VB is given by the terms
$c_pN_2p$ where $c_p$ is
a constant.
We thus require that the
dissipator's structure is such that the master
equation reduces to Eqs. (\ref{mag:eq3})-(\ref{mag:eq2}) when the HFI is lifted ($A=0$).
This may be achieved
by identifying 
$n$, $p$, $N_1$, $N_2$, $\boldsymbol{S}$ and $\boldsymbol{S}_c$
with the quantum statistical average of the corresponding operators, namely
$n=\mathrm{Tr}[\hat\rho\hat{n}]$, $p=\mathrm{Tr}[\hat\rho\hat{p}]$,
$N_1=\mathrm{Tr}[\hat\rho\hat{N_1}]$, $N_2=\mathrm{Tr}[\hat\rho\hat{N}_2]$,
 $\boldsymbol{S}=\mathrm{Tr}[\hat\rho\hat{\boldsymbol{S}}]$ and
 $\boldsymbol{S}_c=\mathrm{Tr}[\hat\rho\hat{\boldsymbol{S}}_c]$.
 As well, the quantum statistical average of
 any generator of $\Lambda$ is given by
 $\lambda_q=\mathrm{Tr}[\hat\rho\hat{\lambda}_q]$
 and therefore, the density matrix can be expanded as
 \begin{eqnarray}\label{ope-densidad}
   \hat{\rho}&=&\sum_{q=1}^{85}
   \frac{\mathrm{Tr}\left[\hat{\rho} \hat{\lambda}_{q}\right]}
{\mathrm{Tr}\left[\hat{\lambda}^{2}_{q}\right]}
   \hat{\lambda}_{q}=\sum_{q=1}^{85}
   \frac{\lambda_{q}}{\mathrm{Tr}
\left[\hat{\lambda}^{2}_{q}\right]}\hat{\lambda}_{q}.
\end{eqnarray}
 The SDR part of the dissipator $\hat{\mathcal{D}}_{\mathrm{SDR}}$ can
 also be expanded
 in terms of the elements of $\Lambda$ as
\begin{equation}\label{dissipator:SDR}
 \hat{\mathcal{D}}_{\text{SDR}}
 =\sum_{q=1}^{85}
\frac{C\left[\hat \lambda_{q}\right] }{\mathrm{Tr}\left[\hat{\lambda}^{2}_{q}\right]}\hat{\lambda}_{q}
=\sum_{q=1}^{85}
\frac{C_q }{\mathrm{Tr}\left[\hat{\lambda}^{2}_{q}\right]}\hat{\lambda}_{q}.
\end{equation}
To determine the coefficients
$C_q\equiv C[\hat \lambda_q]=\mathrm{Tr}[\hat \lambda_q\hat{\mathcal{D}}_{\mathrm{SDR}}]$ we
insert (\ref{dissipator:SDR}) in the master equation (\ref{masterequation})
and multiply by
$\hat n$, $\hat p$, $\hat N_1$, $\hat N_2$, $\hat{\boldsymbol{S}}$
or $\hat{\boldsymbol{S}}_c$.
By taking the trace of the resulting equation
we readily find the coefficients
\begin{eqnarray}
C\left[\hat p\right] &=&-c_p p N_2,\label{bal:1}\\
C\left[\hat n\right] &=&
-c_n\left(nN_1-4\boldsymbol{S}\cdot\boldsymbol{S}_c\right),\\
C\left[\hat N_1\right] &=&
  -c_n\left(nN_1-4\boldsymbol{S}\cdot\boldsymbol{S}_c\right)
  +c_pN_2p,\\
C\left[\hat N_2\right] &=& 
  c_n\left(nN_1-4\boldsymbol{S}\cdot\boldsymbol{S}_c\right)
  -c_pN_2p,\\ 
C\left[\hat{\boldsymbol{S}}\right] &=&
  -c_n\left(\boldsymbol{S}N_1-\boldsymbol{S}_cn\right),\\
C\left[\hat{\boldsymbol{S}}_c\right] &=& 
-c_n\left(\boldsymbol{S_c}n-\boldsymbol{S}N_1\right)\label{bal:6}.
\end{eqnarray}
At this point we have considerable freedom since
these equations only define 10 of the 85 coefficients needed to
fully determine the
$\hat{\mathcal{D}}_{\mathrm{SDR}}$ dissipator.
However, the choices get narrowed down
by imposing the symmetry and invariance properties 
that the system is expected to satisfy.
As the most basic requirement, the master equation
must be invariant under any arbitrary rotation
in accordance with the
space's isotropy. The tensors  $\hat{T}_{j,i}$ and $\hat{U}_{k,j,i}$
must therefore transform by the corresponding laws. 
The complete set of coefficients is thus given by
\begin{eqnarray}
C\left[\hat p\right]&=&-c_p p T_{0,0},\label{balance:1}\\
C\left[\hat n\right]&=&-c_n \left( S_{0} U_{0,0,0}
     -4 \sum_{r=1}^3 S_{r}U_{r,0,0}\right),\label{balance:2}\\
C\left[\hat S_k\right]&=&-c_n \left( S_{k} U_{0,0,0} -S_{0} U_{k,0,0} \right),\label{balance:3}\\
C\left[\hat U_{0,j,i}\right]&=&c_p p T_{j,i}
  \nonumber\\ &&\,\,\,\,\,\,\,- c_n\left(S_{0} U_{0,j,i}
  -4 \sum_{r=1}^3 S_{r} U_{r,j,i}\right),\label{balance:4}\\
C\left[\hat U_{k,j,i}\right]&=&-c_n \left(S_{0} U_{k,j,i}
- S_{k} U_{0,j,i} \right),\label{balance:5}\\
C\left[\hat T_{j,i} \right]&=& -c_p p T_{j,i}\nonumber\\
  && +\,\,\,\,\,\,\, c_n \left( S_{0} U_{0,j,i} -4 \sum_{r=1}^3 
                          S_{r} U_{r,j,i} \right),\label{balance:6} 
  \end{eqnarray}
where $i,j=0,1,2,3$ and $k=1,2,3$.
The spin precession terms $\boldsymbol{\omega}\times\boldsymbol{S}$
and $\boldsymbol{\Omega}\times\boldsymbol{S}_c$,
absent in Eqs. (\ref{balance:1})-(\ref{balance:6}), are accounted for by the
Zeeman terms in the Hamiltonian (\ref{spinHamiltonian}) since they concern the coherent evolution of the system.

The SDR part of the dissipator also excludes
the spin relaxation of CB and bound electrons.
These terms enter
the dissipator through $\hat{\mathcal{D}}_{\mathrm{S}}$ and
$\hat{\mathcal{D}}_{\mathrm{SC}}$ as
\begin{eqnarray}
\hat{\mathcal{D}}_{\mathrm{S}}
  &=& -\frac{1}{\tau_{s}}\sum_{q=3}^{5}\frac{\lambda_{q}}{\mathrm{Tr}[\lambda_{q}^{2}]}
  \hat{\lambda}_{q}=-\frac{2}{\tau_{s}}\sum_{i=1}^{3}S_{i}\hat{S}_{i},\label{diss:s}\\
\hat{\mathcal{D}}_{\mathrm{SC}}
  &=&
-\frac{2}{\tau_{sc}}\sum_{i=1}^{3}S_{ci}\hat{S}_{ci}
.\label{diss:sc}
\end{eqnarray}
The CB electron spin relaxation time due to the Dyakonov-Perel mechanism is given
by $\tau_{s}$, while $\tau_{sc}$ is the phenomenological bound
electron spin relaxation time in Ga centers
\cite{PhysRevB.90.115205,PhysRevB.91.205202}.
The dissipators (\ref{diss:s}) and (\ref{diss:sc}) yield the CB and
bound electron spin relaxation terms $\boldsymbol{S}/\tau_{s}$ and
$\boldsymbol{S}_{c}/\tau_{sc}$ in the TCM.

\begin{figure}
\includegraphics[angle=0,width=0.5 \textwidth]{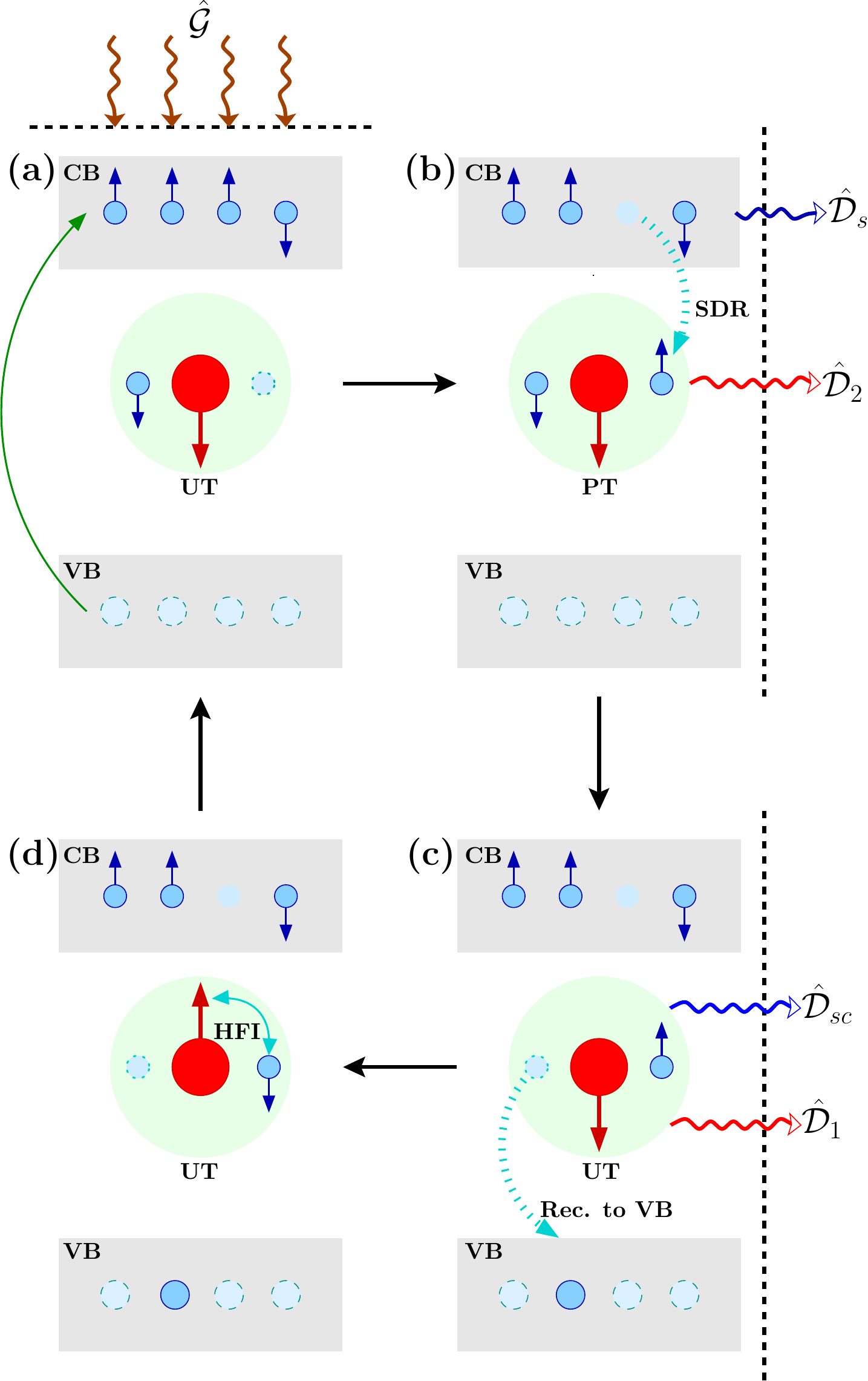}
\caption{
Schematic diagram of the processes involved in the nuclear
spin polarization of Ga centers. The flow of the angular momentum
is also shown. Following the selection rules of GaAs,
three spin up, one spin down CB electrons
and 4 unpolarized VB holes are generated by four
photons with $100\%$ left circular polarization (a).
The angular momentum of photons is transferred to the
CB electrons. 
Traps whose bound electrons are spin
polarized in the opposite direction to the majority of the CB electrons' spin,
can capture CB electrons with opposite spin orientation forming a spin singlet (b).
This process partly transfers angular momentum from the CB electrons to the
bound electrons in traps.
Simultaneously CB electrons' spin and PT's nuclear spins
relax due to $\mathcal{D}_S$ and $\mathcal{D}_2$ making the system
loose angular momentum to the environment.
As one of the trapped electrons recombines to the VB
the spin singlet in the PT is dissociated and becomes a UT (c).
At the same time, the bound electron and nuclear spins in the UT relax
due to the $\mathcal{D}_{\mathrm{SC}}$ and $\mathcal{D}_1$ dissipators  (c).
Again the system looses angular momentum to the environment.
At this stage the bound electron and the nucleus are able to interact  via the
HFI and angular momentum is exchanged (in the presented case) between them by a series ob flip-flops (d).
From (d) to (a) the center can capture a new electron and the cycle starts again.
}\label{figure1}
\end{figure}

\begin{figure}
\includegraphics[angle=0,width=0.48 \textwidth]{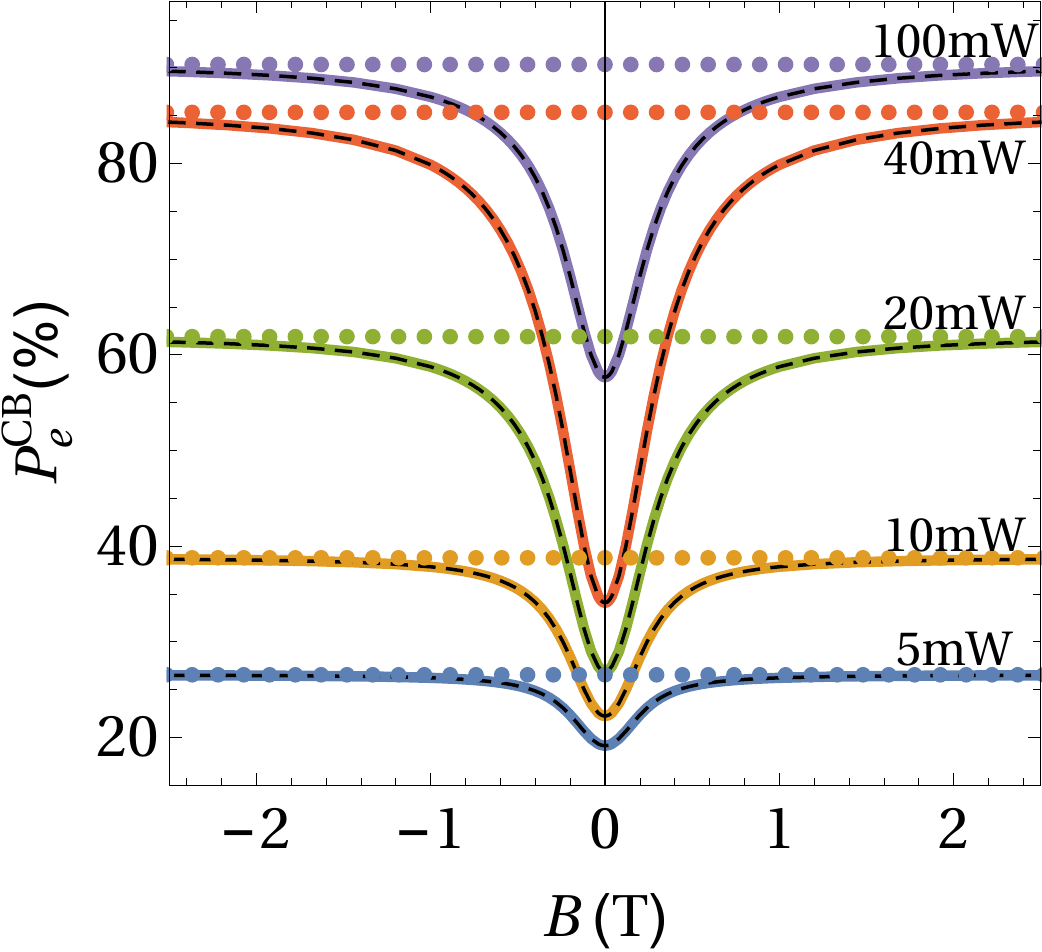}
\caption{
$P^{CB}_e$ as a function of a magnetic field in the Faraday
configuration for various powers $W$ for the non-selective model. 
Results under $\sigma^+$ (solid lines)
and $\sigma^-$ (dashed lines) light excitation are present.
The nuclear spin relaxation times are $\tau_{n1}=5800\, \mathrm{ps}$
 and $\tau_{n2}=533\, \mathrm{ps}$. Dotted horizontal lines present the behaviour of polarization in the absence of NSR.}
\label{figure2}
\end{figure}

\subsection{Nuclear spin relaxation}\label{nsrsec}

To get an insight into the role of the possible mechanisms
involved in NSR
we consider three different models: non-selective, dipolar
and quadrupolar. As a reference
we also study the effects of the absence of spin relaxation.
The dipolar and quadrupolar interactions are dealt through
the Wangsness, Bloch, and Redfield relaxation theory
\cite{ PhysRev.89.728,REDFIELD19651,PhysRev.142.179,kowalewski2006nuclear}
summarized in Appendix \ref{Redfield:theory}. 

First, we study the non-selective\cite{PhysRevB.91.205202}
dissipators for UTs and PTs given by
\begin{eqnarray}
\left(\hat{\mathcal{D}}_1\right)_{sm,s^\prime m^\prime}
&=& -\frac{1}{\tau_{n1}}\bigg(
\rho_{1;s,m;s^\prime, m^\prime}\nonumber\\
&&\,\,
-\frac{\delta_{m,m^{\prime}}}{4}\sum_{m^{\prime\prime}=-3/2}^{3/2}
\rho_{1;s,m^{\prime\prime};s^\prime, m^{\prime\prime}}
\bigg)\:;\\
\left(\hat{\mathcal{D}}_2\right)_{m,m^\prime}
&=& -\frac{1}{\tau_{n2}}\bigg(
\rho_{2;m, m^\prime}\nonumber\\
&&\,\,
-\frac{\delta_{m,m^{\prime}}}{4}\sum_{m^{\prime\prime}=-3/2}^{3/2}
\rho_{2;m^{\prime\prime},m^{\prime\prime}}
\bigg)\:,
\end{eqnarray}
where $s=-1/2,1/2$ and $m=-3/2,-1/2,1/2,3/2$ are the bound
electron spin and nuclear spin indices. This dissipator is highly
symmetrical.
 
Second, we consider the relaxation due to the dipolar
interactions between neighbouring
Ga nuclei.
In this case, the Hamiltonian (\ref{pert:Hamiltonian}) only contains the
irreducible spherical tensors of rank $k=1$.
The terms in this Hamiltonian
correspond to the angular momentum operators interacting
with a random local field.
Substituting (\ref{magdipole:0}) and
(\ref{magdipole:1}) in (\ref{simplified:Redflied})
the dissipators for UTs and PTs are
\begin{eqnarray}
 \hat{\mathcal{D}}_{1}
 &=&-\frac{1}{3\tau_{n1}}\sum_{i=1}^{3}
\left[\hat{I}_{1i},\left[\hat{I}_{1i},
\hat{\rho}\right]\right], \label{D1:D}\\
 \hat{\mathcal{D}}_{2}
 &=&-\frac{1}{3\tau_{n2}}\sum_{i=1}^{3}
\left[\hat{I}_{2i},\left[ 
\hat{I}_{2i},\hat{\rho}\right]\right],\label{D2:D}
\end{eqnarray}
where $\hat{I}_{1i}$ and $\hat{I}_{2i}$ are the
$i$-th components of the nuclear spin
operators' for UTs' and PTs'  respectively.
The NSR times
for unpaired and paired traps
are considered to be different in principle
and therefore are set to $\tau_{n1}$ and $\tau_{n2}$. 

Finally we study the relaxation owing to the quadrupole interaction
with random fluctuation of the local electric field gradient.
The Hamiltonian takes the form
of  (\ref{pert:Hamiltonian}) where $k=2$.
Substituting the rank $k=2$ irreducible spherical tensors
in Eqs. (\ref{elecquadrupole:0})-(\ref{elecquadrupole:2})
in terms of the nuclear angular momentum components in
(\ref{simplified:Redflied}) yield the following
dissipators
\begin{eqnarray}
 \hat{\mathcal{D}}_{1}
 &=&-\frac{1}{2\tau_{n1}}\sum_{i=1}^{5}
\left[\hat{Q}_{1,i},\left[\hat{Q}_{1,i},
\hat{\rho}\right]\right],\label{D1:Q}\\
 \hat{\mathcal{D}}_{2}
 &=&-\frac{1}{2\tau_{n2}}\sum_{i=1}^{5}
\left[\hat{Q}_{2,i},\left[ 
\hat{Q}_{2,i},\hat{\rho}\right]\right].\label{D2:Q}
\end{eqnarray}
Here, the operators $\hat{Q}_{1i}$ and $\hat{Q}_{2i}$
are related to the rank $k=2$ irreducible spherical tensors
and therefore can be expressed in terms of the nuclear
spin operators as
\begin{eqnarray}
\hat{Q}_{n,1}&=& \frac{1}{2\sqrt{3}}\left(\hat{I}^{2}_{nx}
-\hat{I}^{2}_{ny}\right),\\
\hat{Q}_{n,2}&=& \frac{1}{2\sqrt{3}}\left(\hat{I}_{nx}\hat{I}_{ny}
+\hat{I}_{ny}\hat{I}_{nx}\right),\\
\hat{Q}_{n,3}&=&\frac{1}{2\sqrt{3}}\left(\hat{I}_{nx}\hat{I}_{nz}
+\hat{I}_{nz}\hat{I}_{nx}\right),\\
\hat{Q}_{n,4}&=&\frac{1}{2\sqrt{3}}\left(\hat{I}_{ny}\hat{I}_{nz}
+\hat{I}_{nz}\hat{I}_{ny}\right),\\
\hat{Q}_{n,5}&=&\frac{1}{6}\left(2\hat{I}^{2}_{nz}
  -\hat{I}^{2}_{ny}-\hat{I}^{2}_{nx}\right).
\end{eqnarray}
The explicit forms of the dissipators
corresponding to the dipolar interaction
(\ref{D1:D})-(\ref{D2:D}) and quadrupole interaction
(\ref{D1:Q})-(\ref{D2:Q}) are presented
in Appendix \ref{dissipators}.

\section{Results and discussion}\label{resdisc}
The model developed above is used in this section
to examine the interplay of the HFI  and the NSR mechanisms
in the continuous wave (CW) and pulsed excitation (PE) regimes.
First, the theoretical results are compared with previous experimental
observations under CW excitation in order to identify the
main interaction behind NSR and HFI in Ga centers.
Then, we analyze the dynamics of the bound
electrons' and nuclear spin in the PE regime.
We outline a method for detecting the bound electron and
nuclear spin coherent oscillations induced by HFI
by means of a pump-probe PL scheme.

In order to extract information from the model,
we start by building the system of kinetic equations
that follow from the master equation (\ref{masterequation}).  
By multiplying both sides of (\ref{masterequation}) by $\hat{\lambda}_q$,
inserting the density matrix in the form (\ref{ope-densidad}) into
the resulting expression and taking the trace
we obtain a set of $n=85$ differential equations of the form
\begin{multline}
\dot\lambda_q=\frac{i}{\hbar}\mathrm{Tr}\left[\left[\hat H,\hat \lambda_q\right]\hat\rho\right]
+\mathrm{Tr}\left[\mathcal{D}\hat{\lambda}_q\right]\\
=F_q\left(\lambda_1,\lambda_2,\dots,\lambda_n,t\right),
\,\,\,\, q=1,2,\dots, n.\label{kinetic:eqs}
\end{multline}
Unlike the TCM that only considers
the SDR mechanism, Zeeman interaction and electron spin relaxation,
these new kinetic equations also take into account
the HFI and NSR.

We study the spin dynamics of electrons and nuclei by
numerically solving the system of ordinary
differential equations (\ref{kinetic:eqs}).
The relevant parameters are then extracted
from the thus obtained $\lambda_q$ functions which in
turn are the quantum statistical averages. 
We assume that before the optical excitation ($t=0$) the UTs are equally
populated and that the electrons as well as the nuclei are completely
unpolarized. Therefore, initially $N_1(0)=\lambda_6(0)=N_c$
and $\lambda_q=0$ for $q\neq 6$. Notice that these initial
conditions also imply that at this stage
there are no PTs, namely $N_2(0)=\lambda_{70}(0)=0$.

\subsection{Nuclear spin relaxation: CW regime}\label{nsrcwregime}

Under CW stimulation,
the generation of spin up and spin down electrons
is given by the smooth step function
\begin{equation}
G_{\uparrow\downarrow}=
W G\frac{1\pm P_i}{4}\left[1+\tanh\left(\frac{t-t_0}{\sigma}\right)\right],
\end{equation}
where $W$ is the excitation power,
$G=3.0\times 10^{23}\, \mathrm{mW}^{-1}\mathrm{s}^{-1}\mathrm{cm}^{-3}$
is the power to generated electron ratio, $P_i=\pm 0.15$ is the spin
polarization degree of the optically generated
CB electrons, $t_0=100\,\mathrm{ps}$ is the onset time of the excitation
and $\sigma=10\,\mathrm{ps}$ is the duration of the onset.
The system is allowed to evolve for a
sufficiently long time ($200\, \mathrm{ns}$)
to reach steady state conditions.

Some of the parameters as
$N_c=3\times 10^{15}\mathrm{cm}^{-3}$,
$\tau_s=150\,\mathrm{ps}$,
$\tau_{sc}=1500 \,\mathrm{ps}$, $\tau^*=1/c_nN_c=2\,\mathrm{ps}$,
$\tau_h^*=1/c_pN_c=30\,\mathrm{ps}$,
$g=+1$ and $g_c=+2$ where estimated from previous experimental
results
\cite{Kalevich2005,Kalevich2007,PSSA:PSSA200673009,PhysRevB.85.035205}.
For the nuclei at the Ga centers, the hyperfine parameter
was estimated to be
$A=6.9\times 10^{-2}\, \mathrm{cm}^{-1}=8.5\, \mu\mathrm{eV}$,
the average hyperfine parameter of the two stable
isotopes of Ga\cite{:/content/aip/journal/apl/95/24/10.1063/1.3275703,
PhysRevB.85.035205,PhysRevB.90.115205}.
The NSR times $\tau_{n1}$ and $\tau_{n2}$ are determined below
by comparing the theoretical
calculations with the experimental results.

As we stated above,
the two key features behind the HFI in Ga centers
are a growth in the PL degree of circular
polarization $P^{CB}_e$\cite{PhysRevB.87.125202,PhysRevB.91.205202,
PhysRevB.90.115205,PhysRevB.85.035205,Puttisong2013}
and an Overhauser-like magnetic field\cite{Kalevich2013,
PhysRevB.91.205202,PhysRevB.90.115205,
PhysRevB.85.035205,Puttisong2013}.  Both are
observed under circularly polarized excitation
and a Faraday configuration magnetic field.

More specifically
for the first feature,
$P^{CB}_e(B_z)$ exhibits a minimum close to $B_z=0$.
As $|B_z|$ increases, $P^{CB}_e$ saturates at
values above $B_z\approx 25\, \mathrm{mT}$ where Zeeman energies are comparable
to the HFI.
In this region the HFI has been completely
exceeded by the Zeeman interaction and therefore
the bound electrons and nuclei
in the Ga centers are effectively decoupled.
The degree of circular polarization of the CB to VB photoluminescense  
as a function of the Faraday configuration magnetic field
can be described by an inverted Lorentzian-like curve.
In Fig. 2 
we have calculated the CB electron
spin polarization $P^{CB}_e=2 S_z/n$ as a function of
the magnetic field in Faraday configuration $B_z$ under a circularly
polarized excitation. In this case we have chosen
the spin relaxation times $\tau_{n1}=5800\,\mathrm{ps}$ and $\tau_{n2}=533\,\mathrm{ps}$
that give good quantitative agreement with the experiment.
The degree of the CB electron spin polarization
as a function of the Faraday configuration magnetic field
for the non-selective mechanism is shown in Fig. 2 
for different pump powers $W$.
As a reference, Fig. 2 
also
presents the behaviour observed in the absence of NSR
($\mathcal{D}_1=\mathcal{D}_2=0$) as thick dotted lines.
These plots show that despite the HFI,
in the absence of spin relaxation, $P^{CB}_e(B_z)$
does not display any sign of the spin filtering enhancement. 
These results are exactly the same
as those obtained with the TCM that does not
contain the effects of the HFI.
Therefore, in order for the effects of the HFI- as the amplification
of the spin filtering effect-to be visible, NSR is
essential.

\begin{figure}
\includegraphics[angle=0,width=0.485 \textwidth]{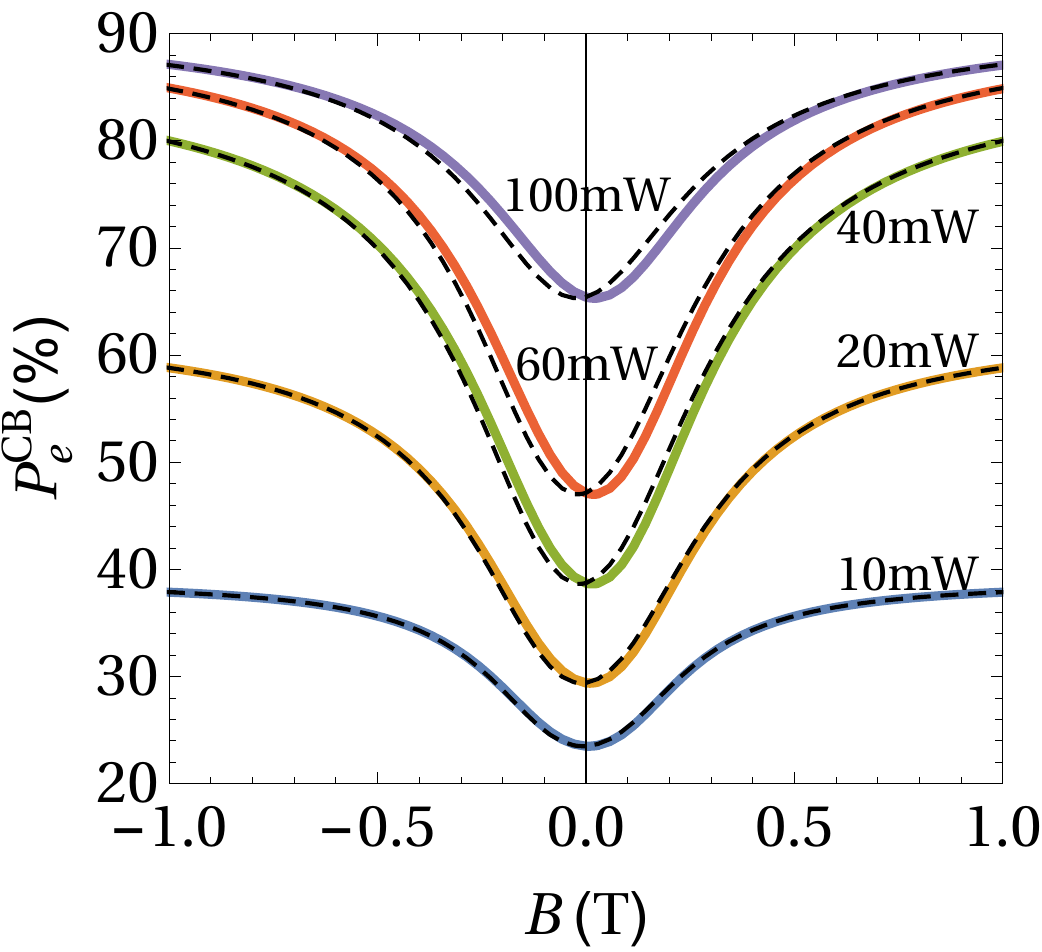}
\caption{
$P^{CB}_e$ as
a function of the magnetic field in the Faraday
configuration for various powers $W$ for the dipolar model. 
Results under $\sigma^+$ (solid lines)
and $\sigma^-$ (dashed lines) light excitation are shown. }
\label{figure3}
\end{figure}

\begin{figure}
\includegraphics[angle=0,width=0.485 \textwidth]{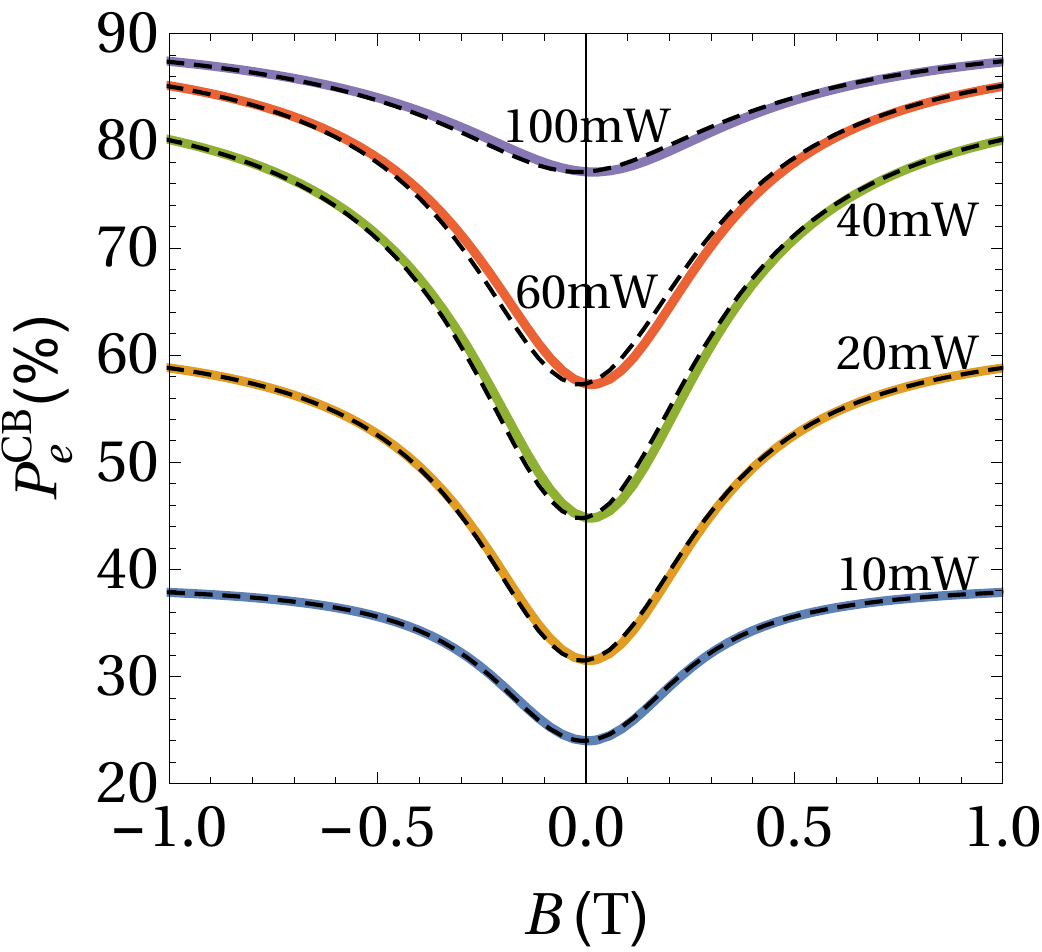}
\caption{
$P^{CB}_e$ as
a function of the magnetic field in the Faraday
configuration for various powers $W$ for the quadrupolar model. 
Results under $\sigma^+$ (solid lines)
and $\sigma^-$ (dashed lines) light excitation are presented.}
\label{figure4}
\end{figure}

\begin{figure}
\includegraphics[angle=0,width=0.48 \textwidth]{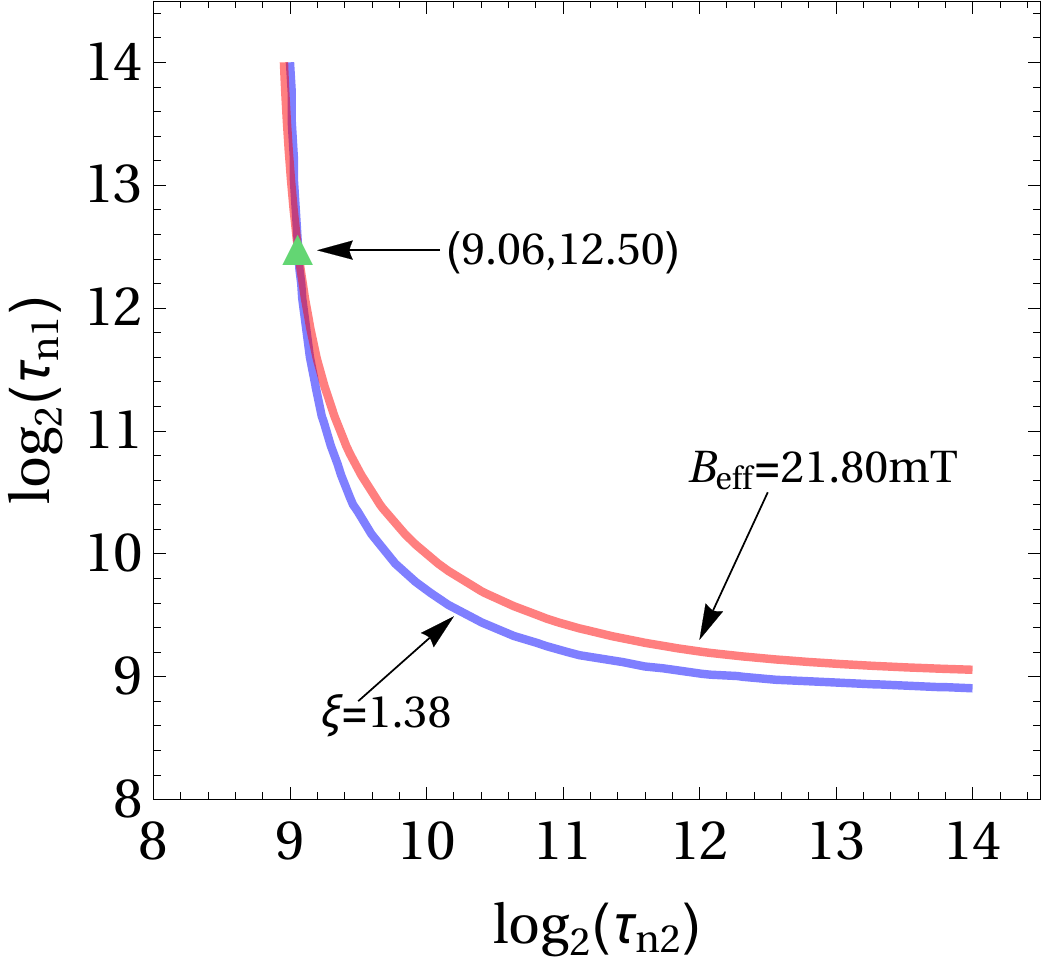}
\caption{
Isolines of the Overhauser-like magnetic field $B_{\mathrm{eff}}$ and depth
$\xi=P^{CB}_e(\infty)/P^{CB}_e(0)$ at fixed power for the
dipolar relaxation mechanism.
The isolines for $B_{\mathrm{eff}}=21.8\, \mathrm{mT}$ and $\xi=1.38$
under an excitation power of $W=100\,\mathrm{mW}$ are shown.
These two lines cross at the point marked with the
(green) triangle in $\tau_{n1}=5800\,\mathrm{ps}$ and $\tau_{n2}=533\,\mathrm{ps}$.
}
\label{figure5}
\end{figure}

\begin{figure}
\includegraphics[angle=0,width=0.23 \textwidth]{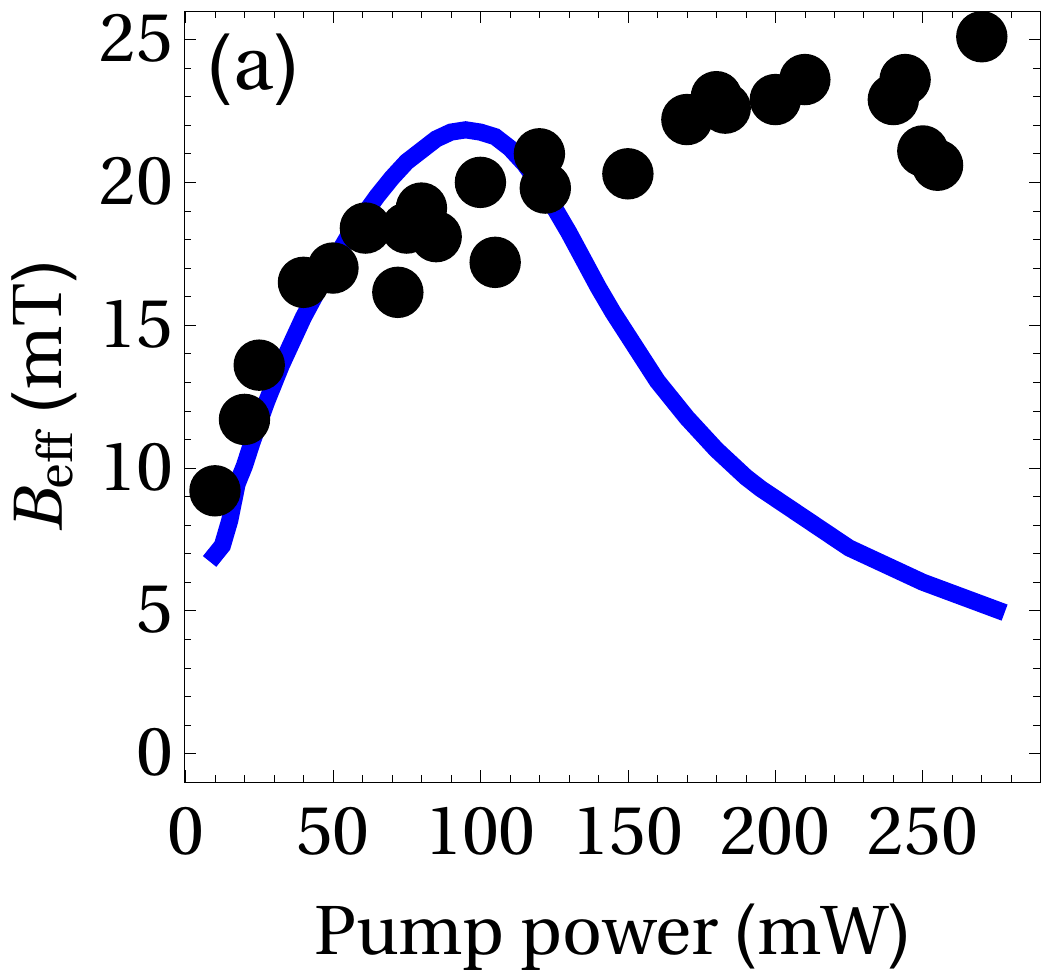}
\includegraphics[angle=0,width=0.23 \textwidth]{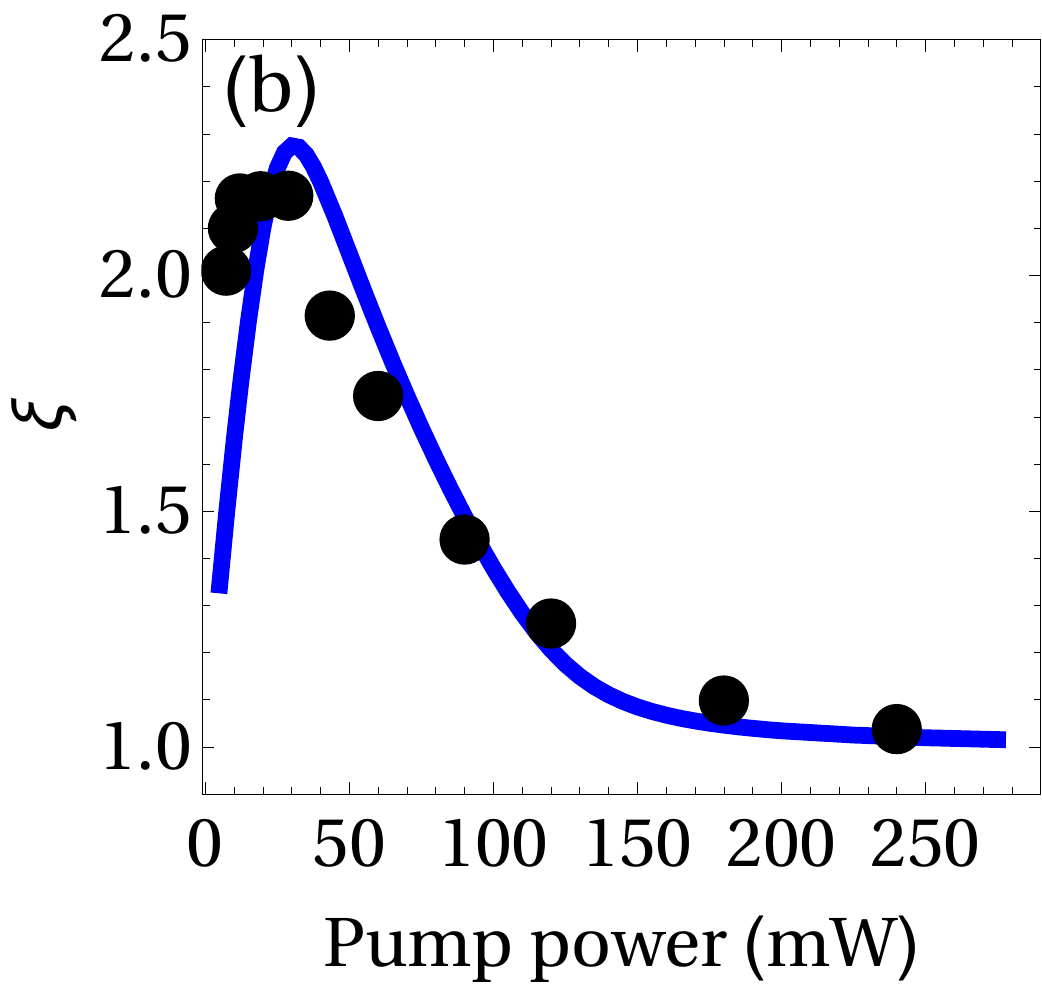}
\caption{
Overhauser-like magnetic field $B_{\mathrm{eff}}$ (a) and $\xi$ (b) as functions of the
excitation power $W$ for the dipolar nuclear spin relaxation mechanism.
In inset (a), the experimental data (solid circles) present the shift
of the intensity dependence $J(B_z)$ while the simulated  curve (solid lines) is obtained for the electron polarization dependence $P^{CB}_e(B_z)$. Inset (b) shows the experimental
(solid circles) \cite{PhysRevB.91.205202,PhysRevB.85.035205}
and theoretical (solid line) results for $\xi$.}
\label{figure6}
\end{figure}

\begin{figure}
\includegraphics[angle=0,width=0.48 \textwidth]{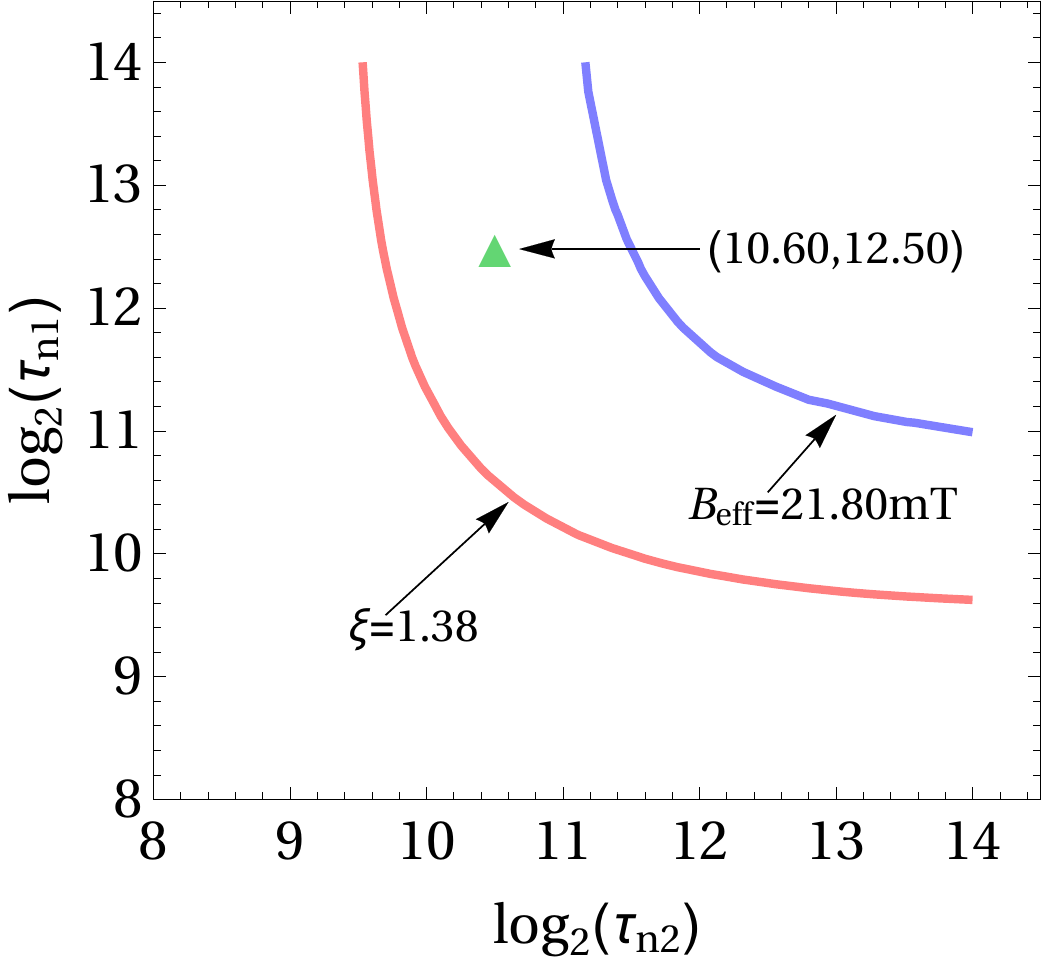}
\caption{
Isolines of the Overhauser-like magnetic field $B_{\mathrm{eff}}$ and depth
$\xi=P^{CB}_e(\infty)/P^{CB}_e(0)$ at fixed power for the
quadrupolar relaxation mechanism.
The isolines for $B_{\mathrm{eff}}=21.80\,\mathrm{mT}$ and $\xi=1.38$
under an excitation power of $W=100\,\mathrm{mW}$ are shown.
Even though these two lines do not cross
we have marked with
a (green) triangle the point $\tau_{n1}=5800\,\mathrm{ps}$
and $\tau_{n2}=1550\,\mathrm{ps}$
that yield the best fit.
}
\label{figure7}
\end{figure}

The second feature of this phenomenon is
a shift of the minimum of $P^{CB}_e$ vs. $B_z$
that points to the existence of an Overhauser-like
magnetic field\cite{Kalevich2013}.
The $P^{CB}_e(B_z)$ curves
are shifted to the positive and negative magnetic
field regions depending on the helicity
of the circularly polarized light.
Thereby, under $\sigma^-$ and $\sigma^+$ excitation
the minimum is located at
$B_z=B_{\mathrm{eff}}<0$ and $B_z=B_{\mathrm{eff}}>0$ respectively.
The experimental data shows that $B_{\mathrm{eff}}$
grows with the excitation power $W$ until it
apparently saturates at approximately $25\, \mathrm{mT}$.
The non-selective dissipator yields vanishing
$B_{\mathrm{eff}}$ as no shift is observed for $P^{CB}_e$ in
Fig. \ref{figure2}.
This dissipator is too symmetric to be able to produce
an Overhauser-like magnetic field and hence 
must be ruled out as the leading NSR mechanism.

In contrast, the dipolar and quadrupolar mechanisms
yield non-vanishing $B_{\mathrm{eff}}$
as it can be seen in Figs. \ref{figure3} 
and \ref{figure4}.
These two plots show $P^{CB}_e(B_z)$ for the dipolar and quadrupolar
dissipators for various excitation powers.
Even though the shifts produced by both mechanisms
qualitatively agree with the experimental observations,
only the dipolar one is able to accurately
fit the experimental measurements as we discuss below.
In addition to the Overhauser-like magnetic field,
another feature that strongly depends on the
NSR mechanism is the depth
of the inverted Lorentzian-like $P^{CB}_e(B_z)$ curves
given by $\xi=P^{CB}_e(\infty)/P^{CB}_e(0)$.
To discern which of the two mechanisms is the
dominant one, we compare
our theoretical calculations with
the experimental
observations of $B_{\mathrm{eff}}$ and $\xi$
\cite{Kalevich2013,PhysRevB.91.205202,PhysRevB.85.035205}.
The power dependence of $B_{\mathrm{eff}}$ and $\xi$
is determined by
finding the minima $P^{CB}_e(0)$ and maxima $P^{CB}_e(\infty)$
of $P^{CB}_e(B_z)$ for different excitation powers.

In Fig. \ref{figure5} 
we plot the isolines
for $B_{\mathrm{eff}}=21.8\,\mathrm{mT}$ and $\xi=1.38$
as functions of the NSR times
$\tau_{n1}$ and $\tau_{n2}$. These two correspond
to the experimental results observed for an excitation power
of $W=100\,\mathrm{mW}$\cite{PhysRevB.85.035205}.
The two isolines intersect at $\tau_{n1}=5800\, \mathrm{ps}$
and $\tau_{n2}=533\, \mathrm{ps}$. In accordance with these results,
the $B_{\mathrm{eff}}$ and $\xi$ isolines at other excitation power
coincide at similar $\tau_{n1}$ and $\tau_{n2}$ values.
Collecting the intersecting points
of all the experimental
results we find that 
the NSR times
must fall within the ranges
$5800\,\mathrm{ps}<\tau_{n1}<8100\,\mathrm{ps}$
and $500\,\mathrm{ps}<\tau_{n2}<700\,\mathrm{ps}$.
Plots of $B_{\mathrm{eff}}$ and $\xi$ as functions of
the excitation power are shown in Figs. \ref{figure6} 
(a) and \ref{figure6}(b)
respectively.
The best quantitative agreement with the experimental data
is accomplished by using the NSR
times $\tau_{n1}=5800 \,\mathrm{ps}$ and
$\tau_{n2}=533\,\mathrm{ps}$ consistent with the ranges above.
Whereas the calculated $\xi$ presents a very good agreement
with the experimental data, the theoretical values of $B_{\mathrm{eff}}$
above $100\,\mathrm{mW}$ show a significant deviation
with respect to the experimental observations.
The experimental results suggest that
after increasing with the excitation power,
$B_{\mathrm{eff}}$ saturates at approximately $25\, \mathrm{mT}$.
However, the computed $B_{\mathrm{eff}}$ as a function
of the excitation vanishes for high powers after
reaching its maximum at $25\, \mathrm{mT}$ (see Fig. \ref{figure6}).

The quadrupolar mechanism, however, yields systematically
non-intersecting isolines regardless of the excitation power
value used to calculate them.
In Fig. \ref{figure7} 
we present
the $B_{\mathrm{eff}}=21.8\, \mathrm{mT}$ and
$\xi=1.38$ isolines
that clearly do not intersect.
This behaviour is observed for all the
excitation powers reported experimentally
and therefore we must also rule out the
quadrupolar mechanism.

\subsection{Coherent oscillations of electronic
and nuclear spins in Ga centers: PE regime}\label{coherent}

\begin{figure}
\includegraphics[angle=0,width=0.48 \textwidth]{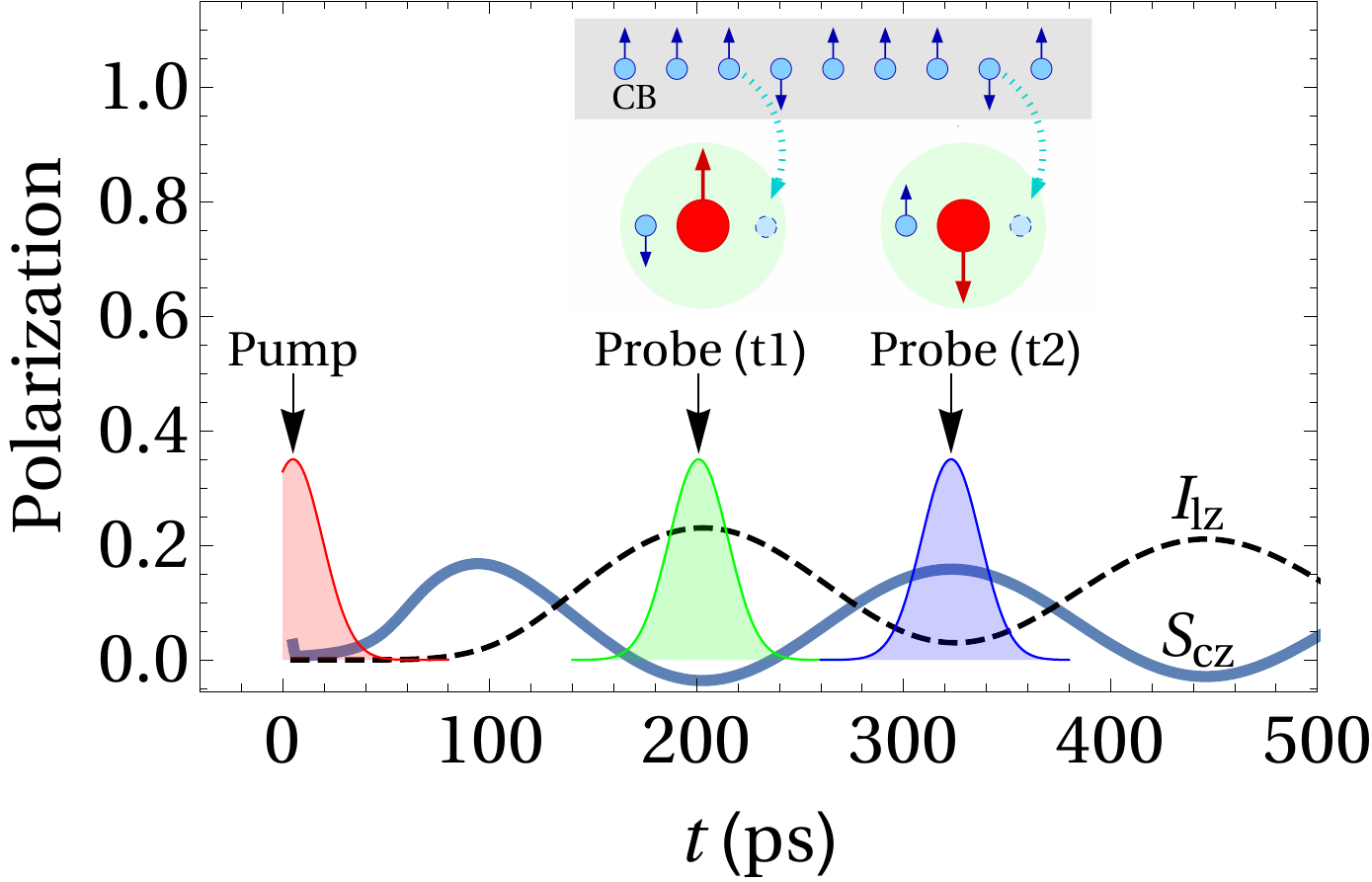}
\caption{
Time dependence of the spin polarization of
bound electrons $S_{cz}$ (solid lines)
and nuclei $I_{1z}$ (dashed lines) after being excited
by the pump pulse.
The pump pulse is left circularly polarized and therefore
the majority of the electrons are spin polarized
in the $+z$ direction.
Two extreme situations are illustrated. In the probe
pulse 1 bound electrons are spin
 polarized in the same
direction as CB electrons and in the probe pulse 2 bound electrons
are spin polarized in the opposite direction to CB electrons.
In the first situation CB electrons with the opposite spin
polarization to the majority are rapidly recombined
through the Ga centers enhancing the spin filtering effect.
In this case a large $SDR_r$ is expected.
In contrast, in the second situation, CB electrons
whose spin polarization is that of the majority are
efficiently recombined lowering the $SDR_r$.
}
\label{figure8}
\end{figure}

\begin{figure}
\includegraphics[angle=0,width=0.48 \textwidth]{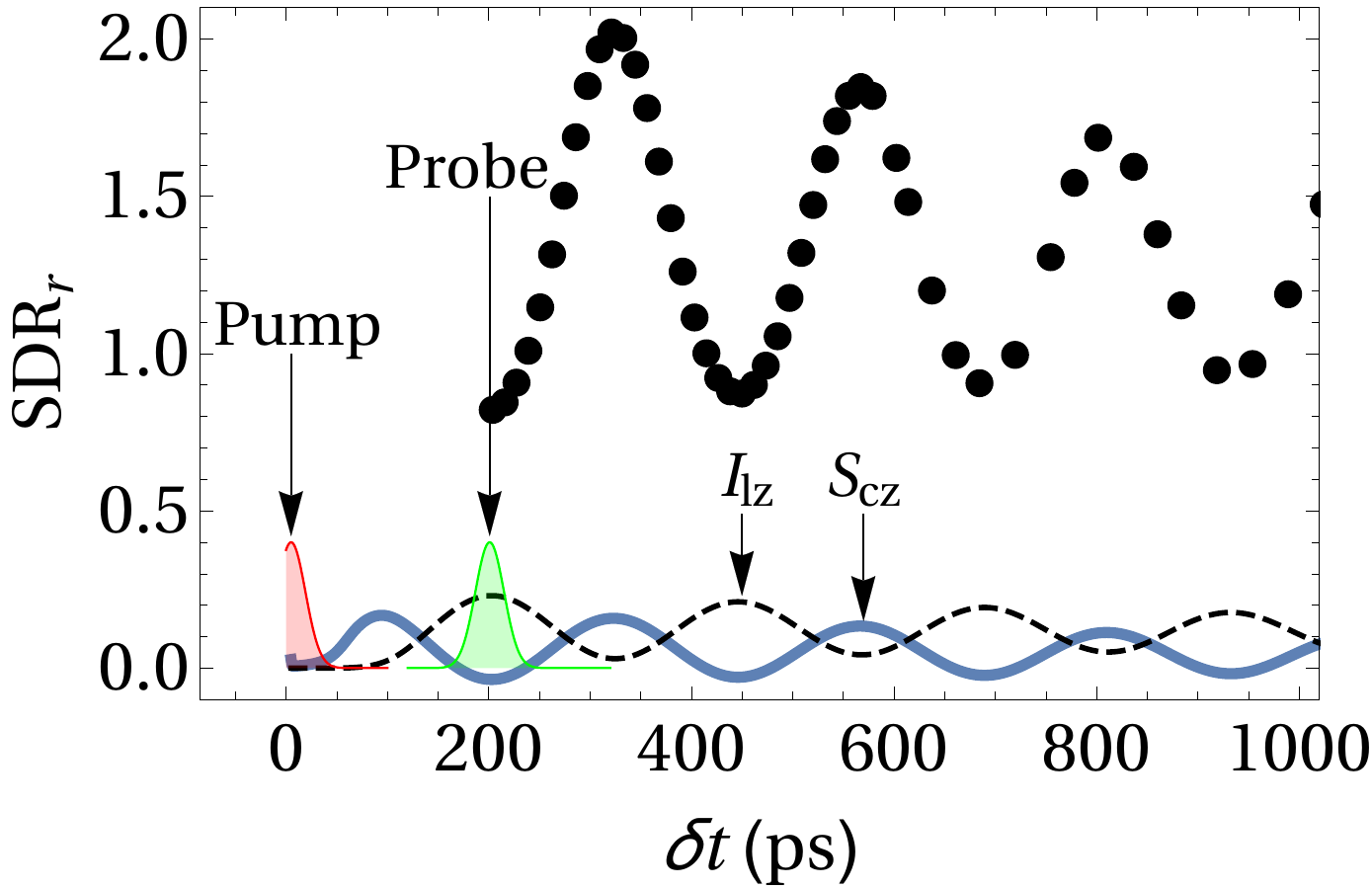}
\caption{
Trace of the coherent electron-nuclear spin oscillations
via the $SDR_r$. The solid circles correspond to maxima
of the time resolved $SDR_r$ as a function of
the time delay $\delta t$ between the pump and the probe pulses.
Below, $S_{cz}$ (solid line) and $I_{1z}$ (dashed line) are
presented for reference.
}
\label{figure9}
\end{figure}

Having identified the NSR mechanism
and the corresponding relaxation times
we are in a position to consider
time resolved simulations.
Our aim here is to develop a method to
observe the coherent oscillations
of bound electrons and nuclei in Ga centers.
To do so we propose a pump probe scheme
that we describe below.

In PE regime, the generation terms are given by
\begin{equation}
G_{\uparrow\downarrow}=
\frac{T W G}{\sigma\sqrt{2\pi}}\frac{1\pm P_i}{2}
\left[
\mathrm{e}^{-\frac{t^2}{2 \sigma^2}}
+\eta\mathrm{e}^{-\frac{(t-\delta t)^2}{2 \sigma^2}}
\right],
\end{equation}
where $W$ is the pulse's average power,
$\sigma=1\, \mathrm{ps}$ is its width
and $T=12$ $\mathrm{ns}$ is the period between
repeated pulses\cite{PhysRevB.83.165202}.
The pump pulse originates at $t=0$
and the probe is delayed $\delta t$.

Figure \ref{figure8} 
is an outline of the proposed method.
The pump pulse is left circularly polarized
and therefore most of the electrons are
spin polarized in the $+z$ direction.
Likewise, the probe pulses are left circularly
polarized
Fig. \ref{figure8} 
 also shows a plot of
the bound electron and nuclear spin polarizations
as a function of time after being excited by the
pump pulse.
The pump and probe pulses widths
are exaggerated to make them visible in the given time scale.
In this plot it is possible to observe the electron-nucleus
flip-flops as oscillations of $I_{1z}$ (dashed lines) and
$S_{cz}$ (thick lines) that are phase shifted by $\pi/2$.
As indicated in this diagram, the second pulse at
time delays $t_1$ and $t_2$ probes two extreme situations.
In the first one CB electron and bound electrons
are mostly spin polarized in opposite directions.
The center is therefore more likely
to capture a CB electron whose spin is oriented in the opposite
direction to the majority therefore rising the population
of electrons in the CB. In contrast, in the second situation
centers are more likely to capture electrons whose direction
is parallel to the majority diminishing the electron population
in the CB.
A good estimate of the electron and hole population in either
situation is the time resolved SDR ratio
given by
\begin{equation}
SDR_r(t)=\frac{I_+}{I_X}=\frac{n_{\sigma_+}(t)p_{\sigma_+}(t)}{n_{\pi_X}(t)p_{\pi_X}(t)}\:,
\end{equation}
where the PL intensity under circularly polarized light
$I_+\propto n_{\sigma_+}(t)p_{\sigma_+}(t)$ is proportional to
the CB and VB density populations $n_{\sigma_+}(t)$ and $ p_{\sigma_+}(t)$.
Similarly $I_X\propto n_{\pi_X}(t)p_{\pi_X}(t)$ where $n_{\pi_X}(t)$ and $p_{\pi_X}(t)$
are the density populations of CB electrons and holes under linearly polarized light.
If CB electrons are captured spin dependently by the Ga centers
then $n_{\sigma_+}(t)>n_{\pi_X}(t))$ and $p_{\sigma_+}(t)>p_{\pi_X}(t)$ and therefore $SDR_r>1$.
In accordance with the above considerations,
$SDR_r(\delta t_1)>SDR_r(\delta t)>SDR_r(\delta t_2)$ where
$\delta t_1<\delta t<\delta t_2$. Thus, it is
possible to trace the oscillations of
bound electrons and nuclei by successively measuring
the time resolved $SDR_r$ for different probe pulse delays.

By determining the maxima of the time
resolved $SDR_r$ for different
probe pulse delays we have obtained
the plot displayed in Fig. \ref{figure9} setting $\eta$=1.
Here the $SDR_r$ maxima are plotted as a function
of their corresponding time delays $\delta t$ as closed circles.
Similar results (not shown here) are obtained by calculating the $SDR_r$
from the integrated PL.
Below, the spin polarization of bound electrons $S_{cz}$
and nuclear spin polarizarion $I_{1z}$ are shown for reference.
This plots demonstrate that it is possible  to trace the
coherent oscillations of the spin polarization
of bound electrons interacting with the nuclei
by means of the time resolved $SDR$ ratio.

\section{Summary }\label{summ}

We have analyzed the spin dynamics of electrons
and nuclei in GaAsN by developing a model
based on the master equation approach.
The main mechanisms behind the spin dependent
recombination are considered as well as
the hyperfine interaction in Ga paramagnetic traps.
We have demonstrated that the NSR
in centers plays an essential role in reproducing the
two most significant signatures of the HFI in Ga centers.
First, the amplification of the spin filtering effect
under a Faraday configuration magnetic field is visible only
if some NSR mechanism is present. Second, the features
of the Overhauser-like magnetic field not only depend on the HFI
but also  strongly rely on the
nature of the NSR mechanism.
We have tested the dipolar interaction between neighbouring
Ga atoms and the quadrupolar interaction of Ga centers
with random charge distribution background.
We have proven that the dipolar is the only mechanism
compatible with the experimental observations.
Indeed, a scenario where large charge distribution variations
are present in the vicinity of the Ga nuclei is difficult
to imagine.
Although most of the experimental results are correctly
reproduced by our model some of the aspects
regarding the behaviour of the Overhauser-like magnetic
field remain elusive. One of these
is the discrepancy between the saturation values
in the high power regime. This is important since
it would allow to pinpoint the exact origin of the NSR mechanism.

To further explore the effects of the HFI and the NSR
we have tested the model in the PE regime.
In particular, we have proposed a pump-probe scheme
that allows to trace the coherent oscillations of
the bound electron spin interacting with its nucleus
through the HFI.

Even though in principle this model is
conceived for Ga centers,
it can be easily adapted for 
other type of centers where dipolar
or quadrupolar interactions play
an important role as the
leading mechanisms of NSR.

\acknowledgments
French and Russian authors 
 acknowledge funding from LIA CNRS-Ioffe RAS ILNACS.
L.A.B., V.K.K. and E.L.I. acknowledge the Russian Foundation for Basic Research (Grant No. 14-02-00959).
A.K. gratefully appreciates the financial support 
of ``Departamento de Ciencias B\'asicas UAM-A" grant numbers
2232214 and 2232215. 
J.C.S.S. and V.G.I.S.
would like to acknowledge the support
received from the
``Becas de Posgrado UAM" scholarship numbers
2151800745 and  2112800069.
We are indebted to Professor J. Grabinsky  
for the careful reading of the manuscript.

\appendix 

\section{Redfield relaxation theory}\label{Redfield:theory}

According to the
Wangsness, Bloch, and Redfield relaxation theory
\cite{ PhysRev.89.728,REDFIELD19651,PhysRev.142.179,kowalewski2006nuclear}
the interaction of a nucleus with its sorroundings
can be accounted for by the Hamiltonian
\begin{equation}\label{pert:Hamiltonian}
 \hat{\mathcal H}\left(t\right)=\gamma\sum_{r=-k}^{k}  f_{k,r}^{*}\left(t\right)\boldsymbol{\hat{T}}_{k,r},
\end{equation}
where $\gamma$ is a constant,  $\boldsymbol{\hat{T}}_{k,r}$
is a $r$-th component of the rank $k$ irreducible spherical 
tensor and $f_{k,r}\left(t\right)$ is a random function
that describes the interaction with the 
surroundings.

To second-order, the average fluctuations of the
surroundings with the nucleus
are given by the following dissipator
\begin{equation}\label{Redflied}
 \hat{\mathcal{D}}
=-\frac{1}{\hbar^2}\int_{-\infty}^{t}dt'
\overline{\left[\hat{\mathcal H}\left(t\right),
\left[\hat{\mathcal H}\left(t'\right),\bar{\hat{\rho}}\right]\right]},
\end{equation}
where $\bar{\hat{\rho}}$ is the average density matrix.
By substituting the Hamiltonian (\ref{pert:Hamiltonian})
in (\ref{Redflied}) 
and using the fact that
$f^{*}_{k,r}\left(t\right)=(-1)^{r}f_{k,-r}\left(t\right)$ and 
$\boldsymbol{\hat{T}}^{\dag}_{k,r}=(-1)^{r}\boldsymbol{\hat{T}}_{k,-r}$
we get the general form of the dissipator
\begin{multline}
\hat{\mathcal{D}}=
-\frac{1}{\hbar^2}\sum_{s,r=-k}^{k}\left[\boldsymbol{\hat{T}}^{\dag}_{k,s},\left[ 
\boldsymbol{\hat{T}}_{k,r} ,\bar{\hat{\rho}}\right]\right]\\
\times
\left(\int_{-\infty}^{t}dt'\gamma^{2}
\overline{f_{k,s}\left(t\right)f_{k,r}^{*}\left(t'\right)}\right).
\label{gen:Redflied}
\end{multline}
The above dissipator  describes the interaction of
a nucleus with the fluctuations of a random electromagnetic field.
The functions $f_{k,s}\left(t\right)$ and $f_{k,r}^{*}\left(t'\right)$
comply with
\begin{equation}\label{condition:fi}
\gamma^{2}\overline{f_{k,s}
\left(t\right)f_{k,r}^{*}\left(t'\right)}
=\delta_{s,r}\xi e^{-\left|t-t'\right|/\tau},
\end{equation}
where $\xi e^{-\left|t-t'\right|/\tau}$ is a correlation function. $\tau$ is the correlation time of the fluctuating field and $\xi$ is 
the correlation amplitude when $t=t'$. 
With (\ref{condition:fi}), the dissipator $\hat{\mathcal{D}}$
 (\ref{gen:Redflied})  is simplified to
\begin{equation}\label{simplified:Redflied}
 \hat{\mathcal{D}}_{\text{\tiny NSR}}=
 -\frac{1}{2\tau_{n}}\sum_{r=-k}^{k}\left[\boldsymbol{\hat{T}}^{\dag}_{k,r},\left[ 
 \boldsymbol{\hat{T}}_{k,r} ,\bar{\hat{\rho}}\right]\right],
\end{equation}
where $\tau_{n}=\hbar^{2}/2\xi\tau$
is the NSR time.

In the case of magnetic dipole interactions, only the
irreducible spherical tensors of rank $k=1$ participate
in the Hamiltonian (\ref{pert:Hamiltonian}).
They can be expressed in terms of the
nuclear spin components as
\begin{eqnarray}
\boldsymbol{\hat{T}}_{1,0}=&&\hat{I}_{z}\label{magdipole:0},\\
\boldsymbol{\hat{T}}_{1,\pm 1}=&&
\mp \frac{1}{\sqrt{2}}\left(\hat{I}_{x}\pm i\hat{I}_{y}\right)\label{magdipole:1},
\end{eqnarray}
and for electric quadrupole interaction the rank is $k=2$ and their components are
\begin{eqnarray}\label{elec:quadrupole}
 \boldsymbol{\hat{T}}_{2,0}=
 &&\frac{1}{6}\left[2\hat{I}^{2}_{z}
-\hat{I}^{2}_{y}-\hat{I}^{2}_{x}\right],\label{elecquadrupole:0}\\
 \boldsymbol{\hat{T}}_{2,\pm1}=&&\mp\frac{1}{2\sqrt{6}}
\left[\hat{I}_{x}\hat{I}_{z}+\hat{I}_{z}\hat{I}_{x}
\pm i\left(
\hat{I}_{y}\hat{I}_{z}+\hat{I}_{z}\hat{I}_{y}\right)
\right]\label{elecquadrupole:1},\\
\boldsymbol{\hat{T}}_{2,\pm 2}=&&
\frac{1}{2\sqrt{6}}\left[\left(\hat{I}^{2}_{x}-\hat{I}^{2}_{y}\right)
\pm i\left(\hat{I}_{x}\hat{I}_{y}+\hat{I}_{y}\hat{I}_{x}\right)
\right]\label{elecquadrupole:2}.
\end{eqnarray}

\section{Explicit form of the dipolar and
 quadrupolar interactions}\label{dissipators}

The quadrupolar dissipator for PT is given explicitly
by
\begin{eqnarray}
&&\left( \hat{\cal D}_2\right)_{\frac32,\frac32} = - \frac{1}{\tau_{n2}} \left( \rho_{2;\frac32, \frac32} - \frac{\rho_{2;\frac12, \frac12} + \rho_{2;-\frac12, -\frac12} }{2} \right)\:,\label{A1}\\
&&\left( \hat{\cal D}_2\right)_{-\frac32,-\frac32} = - \frac{1}{\tau_{n2}} \left( \rho_{2;-\frac32,-\frac32} -\frac{\rho_{2;\frac12, \frac12} + \rho_{2;-\frac12, -\frac12} }{2}\right)\:,\nonumber \\ \label{A2}\\
&& \left( \hat{\cal D}_2\right)_{\frac12,\frac12} = - \frac{1}{\tau_{n2}} \left( \rho_{2;\frac12,\frac12} - \frac{ \rho_{2;\frac32, \frac32} + \rho_{2;-\frac32, -\frac32} }{2} \right)\:,\label{A3}\\
&&\left( \hat{\cal D}_2\right)_{-\frac12,-\frac12}= - \frac{1}{\tau_{n2}} \left(\rho_{2;-\frac12,-\frac12} - \frac{ \rho_{2;\frac32, \frac32} + \rho_{2;-\frac32, -\frac32} }{2} \right)\:,\nonumber \\ \label{A4}\\
&& \left( \hat{\cal D}_2\right)_{\frac32,\frac12} = -\frac{1}{2 \tau_{n2}} \left( 3 \rho_{2;\frac32,\frac12} - \rho_{2;-\frac12,-\frac32} \right)\:, \\
&&\left( \hat{\cal D}_2\right)_{-\frac32,-\frac12} = -\frac{1}{2\tau_{n2}} \left( 3 \rho_{2;-\frac32,-\frac12} - \rho_{2;\frac12,\frac32} \right)\:,\\
&&\left( \hat{\cal D}_2\right)_{-\frac12,-\frac32} = - \frac{1}{2\tau_{n2}} \left( 3 \rho_{2;-\frac12,-\frac32} - \rho_{2;\frac32,\frac12} \right)\:,\label{A5}\\
&&\left( \hat{\cal D}_2\right)_{\frac12,\frac32} = -\frac{1}{2\tau_{n2}} \left( 3 \rho_{2;\frac12,\frac32} - \rho_{2;-\frac32,-\frac12} \right),\label{A6}\\
&&\left( \hat{\cal D}_2\right)_{\frac12,-\frac12} = -\frac{1}{\tau_{n2}} \rho_{2;\frac12,-\frac12} \:,\\
&& \left( \hat{\cal D}_2\right)_{-\frac12,\frac12} = -\frac{1}{\tau_{n2}} \rho_{2;-\frac12,\frac12} \:,\label{A7}
\end{eqnarray}
\begin{eqnarray}
&& \left( \hat{\cal D}_2\right)_{\frac32,-\frac12} = -\frac{1}{2 \tau_{n2}} \left( 3 \rho_{2;\frac32,-\frac12} + \rho_{2;\frac12,-\frac32} \right)\:,\\
&&
\left( \hat{\cal D}_2\right)_{\frac12,-\frac32} = - \frac{1}{2\tau_{n2}} \left( 3 \rho_{2;\frac12,-\frac32} + \rho_{2;\frac32,-\frac12} \right)\:,\label{A8}\\
&& \left( \hat{\cal D}_2\right)_{-\frac32,\frac12} = -\frac{1}{2\tau_{n2}} \left( 3 \rho_{2;-\frac32,\frac12} + \rho_{2;-\frac12,\frac32} \right)\:, \\&&
\left( \hat{\cal D}_2\right)_{-\frac12,\frac32} = - \frac{1}{2\tau_{n2}} \left( 3 \rho_{2;-\frac12,\frac32} + \rho_{2;-\frac32,\frac12} \right),\label{A9}\\
&& \left( \hat{\cal D}_2\right)_{\frac32,-\frac32} = -\frac{1}{\tau_{n2}} \rho_{2;\frac32,-\frac32} \:,\\&&\left( \hat{\cal D}_2\right)_{-\frac32,\frac32} = -\frac{1}{\tau_{n2}} \rho_{2;-\frac32,\frac32}\: .
\label{A10}
\end{eqnarray}
In a short form the above equations can be presented as
\[
\left( \hat{\cal D}_2\right)_{m,m'} = -\frac{1}{\tau_{n2}}\sum_{m_1,m'_1} Q^{({\rm EQ})}_{m,m';m_1,m'_1} \rho_{2; m_1,m'_1}\:.
\] 
Then the quadrupolar dissipator for UT is given by
\[
\left( \hat{\cal D}_1\right)_{s,m;s',m'} = -\frac{1}{\tau_{n1}}\sum_{m_1,m'_1}Q^{({\rm EQ})}_{m,m';m_1,m'_1} \rho_{1; s,m_1;s',m'_1}\:.
\] 

The dipolar dissipator for PTs is given explicitly by
\begin{eqnarray} 
&&\left( \hat{\cal D}_2\right)_{\frac32,\frac32} = - \frac{1}{\tau_{n2}} \left( \rho_{2;\frac32,\frac32} - \rho_{2;\frac12,\frac12}\right)\:,\\
&& \left( \hat{\cal D}_2\right)_{-\frac32,-\frac32} = - \frac{1}{\tau_{n2}} \left( \rho_{2;-\frac32,-\frac32} - \rho_{2;-\frac12,-\frac12}\right),\\
&& \left( \hat{\cal D}_2\right)_{\frac12,\frac12} = - \frac{1}{3\tau_{n2}} \left(7 \rho_{2;\frac12,\frac12} - 3 \rho_{2;\frac32,\frac32} - 4 \rho_{2;-\frac12,-\frac12}\right),\\
&&\left( \hat{\cal D}_2\right)_{-\frac12,-\frac12}= - \frac{1}{3\tau_{n2}} \left(7 \rho_{2;-\frac12,-\frac12} - 3 \rho_{2;-\frac32,-\frac32} - 4 \rho_{2;\frac12,\frac12}\right),\nonumber \\ \\
&& \left( \hat{\cal D}_2\right)_{\frac32,\frac12} = - \frac{2}{3\tau_{n2}} \left( 3 \rho_{2;\frac32,\frac12} - \sqrt{3} \rho_{2;\frac12,-\frac12}\right),  \\
&&\left( \hat{\cal D}_2\right)_{\frac12,-\frac12} = - \frac{2}{3\tau_{n2}} \left[ 4 \rho_{2;\frac12,-\frac12} - \sqrt{3} \left( \rho_{2;\frac32,\frac12} + \rho_{2;-\frac12,-\frac32} \right) \right]\:,\nonumber \\ \\
&&\left( \hat{\cal D}_2\right)_{-\frac12,-\frac32} = - \frac{2}{3\tau_{n2}} \left( 3 \rho_{2;-\frac12,-\frac32} - \sqrt{3} \rho_{2;\frac12,-\frac12}\right),  \\
&& \left( \hat{\cal D}_2\right)_{\frac12,\frac32} = - \frac{2}{3\tau_{n2}} \left( 3 \rho_{2;\frac12,\frac32} - \sqrt{3} \rho_{2;-\frac12,\frac12}\right), \\
&&\left( \hat{\cal D}_2\right)_{-\frac12,\frac12} = - \frac{2}{3\tau_{n2}} \left[ 4 \rho_{2;-\frac12,\frac12} - \sqrt{3} \left( \rho_{2;\frac12,\frac32} + \rho_{2;-\frac32,-\frac12} \right) \right],\nonumber \\ \\
&&\left( \hat{\cal D}_2\right)_{-\frac32,-\frac12} = - \frac{2}{3\tau_{n2}} \left( 3 \rho_{2;-\frac32,-\frac12} - \sqrt{3} \rho_{2;-\frac12,\frac12}\right), \\
&&\left( \hat{\cal D}_2\right)_{\frac32,-\frac12} = - \frac{1}{\tau_{n2}} \left( 3 \rho_{2;\frac32,-\frac12} - \rho_{2;\frac12,-\frac32} \right)\:,\\ &&\left( \hat{\cal D}_2\right)_{\frac12,-\frac32} = - \frac{1}{\tau_{n2}} \left( 3 \rho_{2;\frac12,-\frac32} - \rho_{2;\frac32,-\frac12} \right),\\
&&\left( \hat{\cal D}_2\right)_{-\frac32,\frac12} = - \frac{1}{\tau_{n2}} \left( 3 \rho_{2;-\frac32,\frac12} - \rho_{2;-\frac12,\frac32} \right)\:,\\&&
\left( \hat{\cal D}_2\right)_{-\frac12,\frac32} = - \frac{1}{\tau_{n2}} \left( 3 \rho_{2;-\frac12,\frac32} - \rho_{2;-\frac32,\frac12} \right),\\
&&\left( \hat{\cal D}_2\right)_{\frac32,-\frac32} = - \frac{4}{\tau_{n2}} \rho_{2;\frac32,-\frac32}\:,\\&&\left( \hat{\cal D}_2\right)_{-\frac32,\frac32} = - \frac{4}{\tau_{n2}} \rho_{2;-\frac32,\frac32} \:.
\end{eqnarray}

If the above equations are presented as
\[
\left( \hat{\cal D}_2\right)_{m,m'} = -\frac{1}{\tau_{n2}}\sum_{m_1,m'_1}
Q^{({\rm MD})}_{m,m';m_1,m'_1} \rho_{2; m_1,m'_1}\:,
\] 
then the depolar dissipator for UT is given by
\[
\left( \hat{\cal D}_1\right)_{s,m;s',m'} = -\frac{1}{\tau_{n1}}\sum_{m_1,m'_1}Q^{({\rm MD})}_{m,m';m_1,m'_1} \rho_{1; s,m_1;s',m'_1}\:.
\]

%

\end{document}